\documentclass[aps,pra,twocolumn,showpacs,superscriptaddress,groupedaddress]{revtex4-1} 
\usepackage{graphicx,latexsym}
\usepackage{amsmath,amssymb,amsfonts}
\usepackage{bm}
\usepackage{braket}
\usepackage{mathrsfs}
\usepackage{color}
\usepackage[normalem]{ulem}

\newcommand{\up}{\uparrow}
\newcommand{\down}{\downarrow}
\newcommand{\m}{\mathcal}
\newcommand{\ep}{\varepsilon}
\def\diamondcomma{\ \raise.3ex\hbox{$\diamond$}\kern-0.4em\lower.7ex\hbox{$,$}\ }
\def\lesssim{\ \raise.3ex\hbox{$<$}\kern-0.8em\lower.7ex\hbox{$\sim$}\ }
\def\gesim{\ \raise.3ex\hbox{$>$}\kern-0.8em\lower.7ex\hbox{$\sim$}\ }

\begin{document}
\title{
Proposed Fermi-surface reservoir-engineering and application to realizing unconventional Fermi superfluids in a driven-dissipative non-equilibrium Fermi gas}
\author{Taira Kawamura\email{tairakawa@keio.jp}}
\affiliation{Department of Physics, Keio University, 3-14-1 Hiyoshi, Kohoku-ku, Yokohama 223-8522, Japan}
\author{Ryo Hanai}
\affiliation{Asia Pacific Center for Theoretical Physics, Pohang 37673, Korea}
\affiliation{Department of Physics, POSTECH, Pohang 37673, Korea}
\author{Yoji Ohashi}
\affiliation{Department of Physics, Keio University, 3-14-1 Hiyoshi, Kohoku-ku, Yokohama 223-8522, Japan}
\date{\today}
\begin{abstract}
We develop a theory to describe the dynamics of a driven-dissipative many-body Fermi system, to pursue our proposal to realize exotic quantum states based on reservoir engineering. Our idea is to design the shape of a Fermi surface so as to have multiple Fermi edges, by properly attaching multiple reservoirs with different chemical potentials to a fermionic system. These emerged edges give rise to additional scattering channels that can destabilize the system into unconventional states, which is exemplified in this work by considering a driven-dissipative attractively interacting Fermi gas. By formulating a quantum kinetic equation using the Nambu-Keldysh Green's function technique, we explore nonequilibrium steady states in this system and assess their stability. We find that, in addition to the BCS-type isotropic pairing state, a Fulde-Ferrell-type anisotropic superfluid state being accompanied by Cooper pairs with non-zero center-of-mass momentum exists as a stable solution, even in the absence of a magnetic Zeeman field. Our result implies a great potential of realizing quantum matter beyond the equilibrium paradigm, by engineering the shape and topology of Fermi surfaces in both electronic and atomic systems. 
\end{abstract}
\maketitle
%%%%%%%%%%%%%%%%%%%%%%%%%%%%%%%%%%%%%%%%%%%%%%%%%%%%%%%%%%%%%%%%%%%%%%%%%%
\newpage
\section{Introduction}
\par
The last two decades had witnessed great progress in understanding and controlling many-body systems out of equilibrium \cite{Marchetti2013, Carusotto2013, Sieberer2016, Ashida2020, Bukov2015, Eckardt2017, Oka2019, Harper2020, Aoki2014}. When the system is driven out of equilibrium, restrictions such as the fluctuation-dissipation theorem are generally lifted. The lack of these constraints gives additional ``free hands'' for the system to exhibit exotic states that are otherwise prohibited in equilibrium. Floquet time crystals \cite{Khemani2016, Else2016, Yao2017, Else2017, Zhang2017, Choi2017}, light-induced superconducting-like states \cite{Fausti2011, Mitrano2016, Suzuki2019}, long-range orders in two dimensions \cite{Vicsek1995, Toner1995} as well as non-reciprocal phase transitions \cite{Fruchart2021, Hanai2020, Hanai2019, You2020, Saha2020}, are a few of such examples. Among these, the strategy of dissipatively controlling many-body states by carefully designing the coupling between reservoirs and a system, which is often referred to as `reservoir engineering', is recognized as a promising route to obtain the desired state \cite{Diehl2008, Verstraete2009, Weimer2010, Diehl2011, Metelmann2015, Ma2019, Clark2020}.  For example, by an appropriate design of reservoir-system coupling, it has been shown to be possible to implement a non-trivial topological state \cite{Diehl2008}, universal quantum computing \cite{Verstraete2009} as well as non-reciprocal coupling  \cite{Metelmann2015}.  Although most of them consider Markovian reservoirs, it has been pointed out that a non-Markovian reservoir is also useful as a dissipative stabilizer of strongly correlated states, such as the Mott insulator \cite{Ma2019} and the fractional quantum Hall state \cite{Clark2020}. 
\par
\begin{figure}[tb]
\centering
\includegraphics[width=7.6cm]{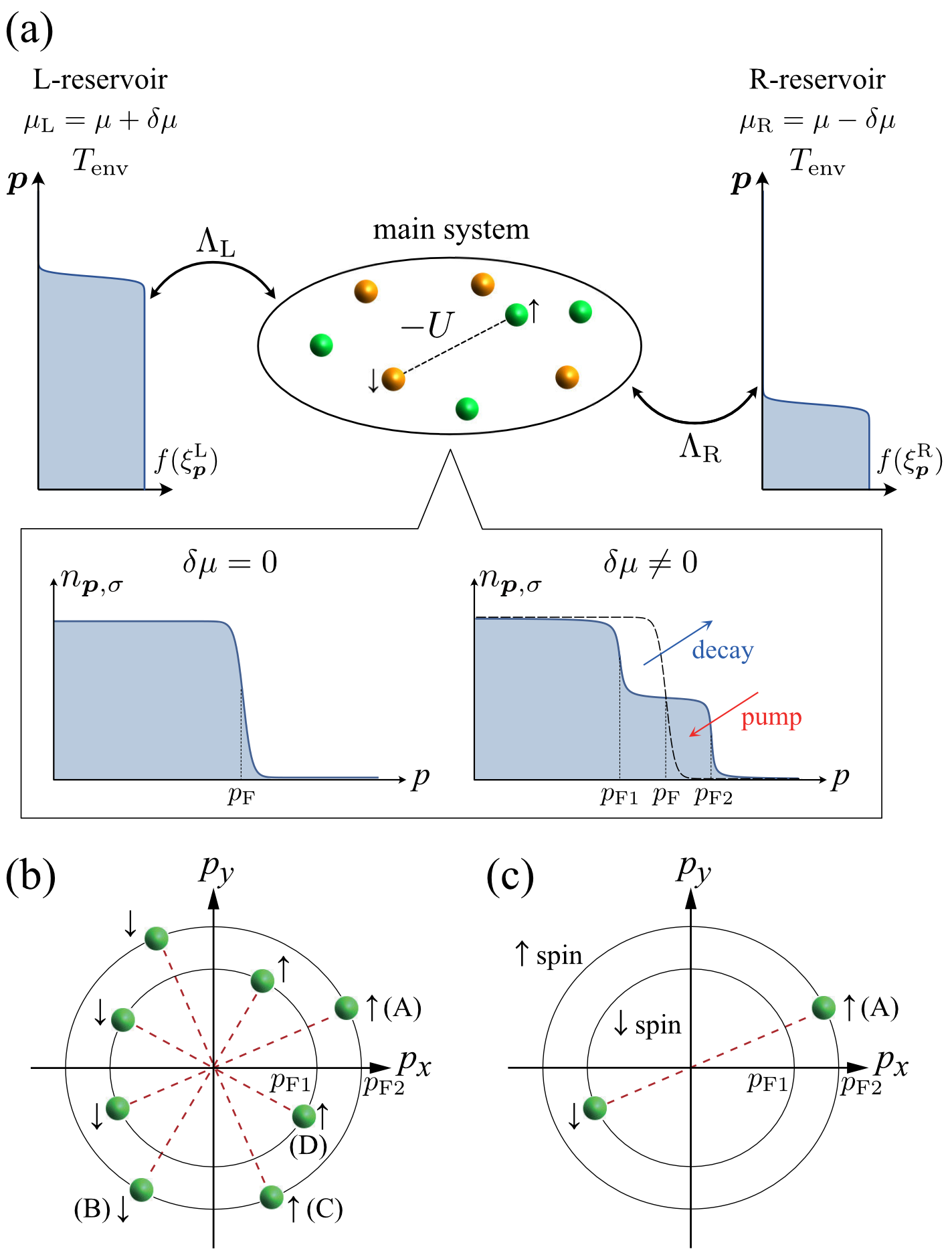}
\caption{(a) Model non-equilibrium driven-dissipative two-component Fermi gas with an $s$-wave pairing interaction $-U (<0)$. The main system is coupled with two reservoirs ($\alpha={\rm L, R}$) with different chemical potentials $\mu_{\rm L}=\mu+\delta\mu$ and $\mu_{\rm R}=\mu-\delta\mu$. Both the reservoirs consist of free fermions in the thermal equilibrium state at the environment temperature $T_{\rm env}$. $f(\xi_{\bm p}^\alpha)$ is the Fermi distribution function, where $\xi_{\bm p}^\alpha=\ep_{\bm{p}}-\omega_\alpha$ is the kinetic energy, measured from $\omega_\alpha$. $\Lambda_{\rm \alpha}$ describes tunneling between the main system and the $\alpha$-reservoir. The pumping and decay of Fermi atoms by the two reservoirs bring about two edges at $p_{\rm F1}=\sqrt{2m\mu_{\rm R}}$ and $p_{\rm F2}=\sqrt{2m\mu_{\rm L}}$ in the Fermi momentum distribution $n_{{\bm p},\sigma}$ in the main system (where $\sigma=\uparrow,\downarrow$ describe two atomic hyperfine states). (b) Expected types (A)-(D) of Cooper pairs, when the two edges work like two Fermi surfaces. (c) Ordinary (thermal equilibrium) Fulde-Ferrell (FF) pairing state under an external magnetic field. 
}
\label{fig1} 
\end{figure}

In this paper, we apply non-Markovian reservoir engineering to a many-body Fermi system. In particular, we propose to design the shape of a Fermi surface by attaching multiple reservoirs with different chemical potentials $\mu_\alpha$ to the main system, to explore exotic quantum many-body states that have not been realized/discussed/known in condensed matter physics. In the simplest model shown in Fig. \ref{fig1}(a), for example, the two reservoirs supply fermion to the system up to their respective chemical potentials, giving rise to a non-equilibrium steady state (NESS) with a two-step structure in the Fermi momentum distribution $n_{{\bm p},\sigma=\up,\down}$. (See the pop-up in Fig. \ref{fig1}(a).) As we discuss in the main text (see also Appendix C), we expect our proposal can be well achieved experimentally in ultracold Fermi gas systems, as well as electron systems, by using the current state-of-art techniques.
\par
%%%%%%%%%%%%%%%%%%%%%%%%%%%%%%%%%%%%%%%%%%%%%%%%%%%%%%%%%%%%%%%%%%%%%%%%%%%%
%%%%%%%%%%%%%%%%%%%%%%%%%%%%%%%%%%%%%%%%%%%%%%%%%%%%%%%%%%%%%%%%%%%%%%%%%%%%
\par
If each edge imprinted on $n_{{\bf p},\sigma}$ works like a Fermi surface, it means that one can produce multiple Fermi surfaces from one Fermi sphere. Then, in the model case in Fig. \ref{fig1}(a), an $s$-wave attractive interaction $-U$ is expected to produce four types of Cooper pairs (A)-(D) shown in Fig. \ref{fig1}(b). Among them, while (C) and (D) are essentially the same as the ordinary BCS pairing, (A) and (B) are unconventional pairings with non-zero center-of-mass momentum. The latter are analogous to the unconventional Fulde-Ferrell (FF) state (see Fig. \ref{fig1}(c)) discussed in superconductivity under an external magnetic field \cite{Fulde1964, Larkin1964, Takada1969, Shimahara1994, Matsuda2007}, spin-polarized Fermi gases \cite{Hu2006, Parish2007, Liao2010, Chevy2010, Kinnunen2018, Strinati2018}, and color superconductivity in quantum chromodynamics \cite{Casalbuoni2004}. We recall that the FF state is usually realized in the spin-imbalanced case, where FF Cooper pairs are formed between $\up$-spin fermions around the larger Fermi surface in Fig. \ref{fig1}(c) and $\down$-spin ones around the smaller Fermi surface, as symbolically written as $|{\rm (A)}\rangle=|-{\bm p}_{\rm F1},\down\rangle|{\bm p}_{\rm F2},\up\rangle$. In contrast, the model driven-dissipative Fermi gas in Fig. \ref{fig1}(a) is {\it not} accompanied by any spin imbalance, but each spin component has two ``Fermi edges" at $p_{\rm F1}$ and $p_{\rm F2}$. This leads to the pairing $|{\rm (B)}\rangle=|{\bm p}_{\rm F1},\up\rangle|-{\bm p}_{\rm F2},\down\rangle$, in addition to $|{\rm (A)}\rangle$. In a sense, the non-equilibrium FF-like (NFF) state may be viewed as a mixture of two FF states under external magnetic fields ${\bm B}$ and $-{\bm B}$.
\par
We note that possible routes to the FF state in the spin-balanced case has been discussed in the literature, where the shift of single-particle energy induced by external current \cite{Doh2006}, a size effect \cite{Vorontsov2009}, an inter-atomic interaction \cite{He2018}, and an artificial field \cite{Zheng2015, Zheng2016, Nocera2017}, have been proposed to realize this unconventional Fermi superfluid. However, these ideas are all in the thermal equilibrium case with the ordinary Fermi distribution function, which is quite different from our idea in the non-equilibrium state.
\par
\begin{figure}[tb]
\centering
\includegraphics[width=7.8cm]{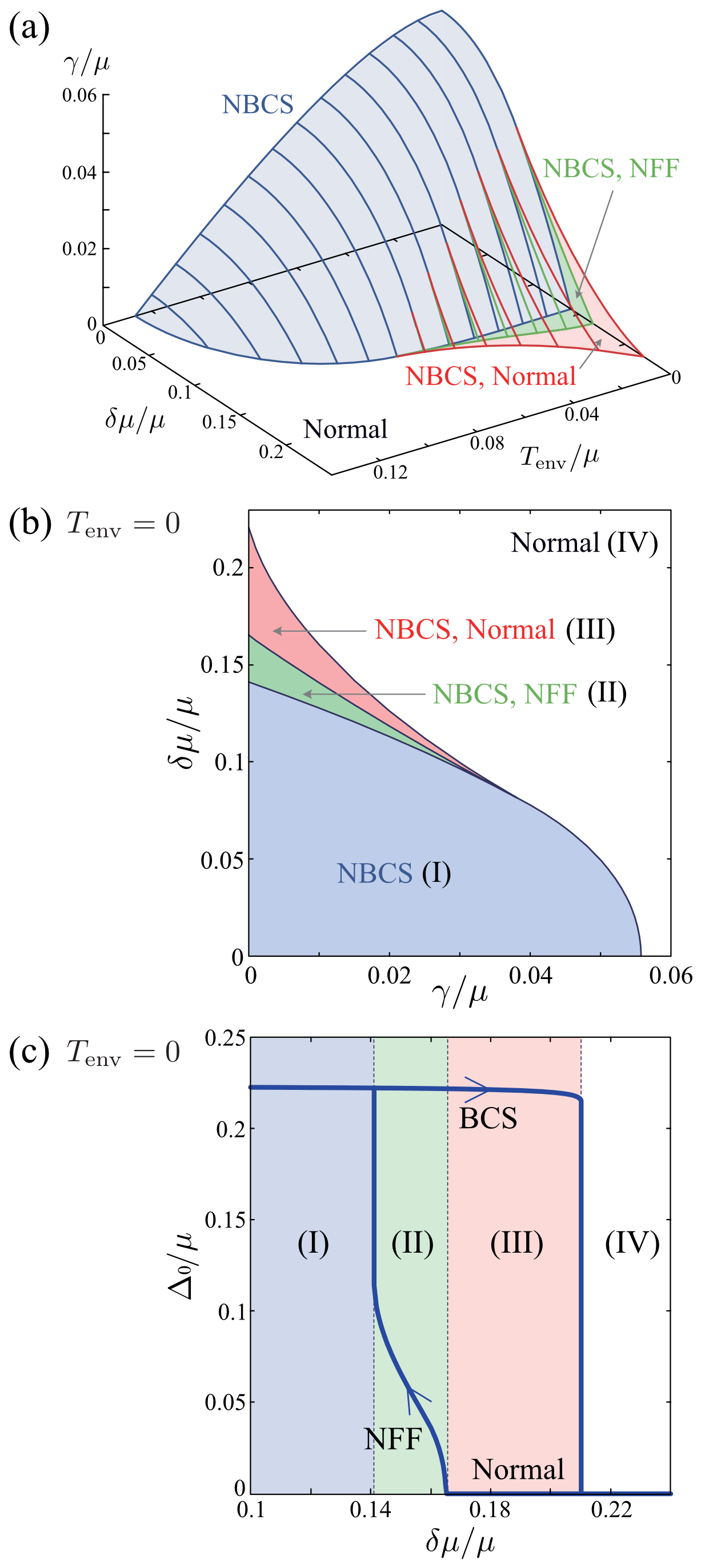}
\caption{Summary of our main results in this paper. (a) Steady-state phase diagram of a driven-dissipative two-component Fermi gas, with respect to half the chemical potential difference $\delta\mu=[\mu_{\rm L}-\mu_{\rm R}]/2$ between the two reservoirs, the damping rate $\gamma$ caused by system-reservoir couplings, and the environment temperature $T_{\rm env}$ (that are all scaled by the averaged chemical potential $\mu=[\mu_{\rm L}+\mu_{\rm R}]/2$). This figure shows the weak-coupling case when $(p_{\rm F}a_s)^{-1}=-1$ (where $p_{\rm F}=\sqrt{2m\mu}$) under the vanishing current condition, ${\bm J}_{\rm net}=0$. In the phase diagram, NBCS and NFF represent the non-equilibrium BCS (${\bm Q}=0$) and FF like (${\bm Q}\ne 0$) states, where ${\bm Q}$ is the center-of-mass momentum of a Cooper pair. (b) The steady-state phase diagram at $T_{\rm env}=0$. (c) Hysteresis phenomenon in the regions (II) and (III), shown in panel (b). In panel (c), we set $T_{\rm env}=0$ and $\gamma\to 0^+$. 
\vspace{-3cm}
}
\label{fig2} 
\end{figure}

In what follows, we confirm our scenario by dealing with the model driven-dissipative Fermi gas in Fig.
\ref{fig1}(a). Our principal results are captured in Fig. \ref{fig2}. Panels (a) and (b) show the steady-state phase diagram of a driven-dissipative Fermi gas, with respect to half the chemical potential difference $\delta\mu=[\mu_{\rm L}-\mu_{\rm R}]/2$ between the two reservoirs, the atomic damping rate $\gamma$ coming from the system-reservoir couplings, and the environment temperature $T_{\rm env}$. We clarify that all the states appearing in Fig. \ref{fig2} are (meta-)stable in the sense that the time evolution of a small {\it deviation} from each state always decays. (Detailed results of the stability analysis are presented in Sec. IV B.) The expected NFF state appears in the region (II), where the chemical potential difference $\delta\mu/\mu$ is large enough to produce a clear two-step structure in $n_{{\bm p},\sigma}$ but the damping $\gamma/\mu$ is not strong enough to smear out this structure. We note that the BCS-type superfluid (NBCS) is also stable in the region (II). This so-called bistability is a characteristic non-equilibrium phenomenon and has been observed in various systems \cite{Labouvie2015, Wang2018, Goldman1987}. This is quite different from the thermal equilibrium case, where the ground state is uniquely identified as the state with the lowest free energy. In the region (III), the bistability of the NBCS and the normal state occurs. 
\par
In the bistability regions (II) and (III), which state (among the multiple meta-stable states) is realized would depend on how the parameters are varied to reach these regions. When one varies $\delta\mu$ adiabatically, we argue that the hysteresis shown in Fig. \ref{fig2}(c) appears: As $\delta\mu$ increases from $\delta\mu=0$, the NBCS would be maintained both in the regions (II) and (III). As one decreases $\delta\mu$ from the region (IV), on the other hand, the phase transition from the normal state to NFF would occur at the boundary between (II) and (III).
\par
To close this section, let us briefly comment on the connection between our previous work \cite{Kawamura2020, Kawamura2020_JLTP} and the present study. In Refs. \cite{Kawamura2020, Kawamura2020_JLTP}, we theoretically studied the properties of a non-equilibrium driven-dissipative Fermi gas in the {\it normal} phase, and found that the chemical potential bias applied by two reservoirs gives rise to the anomalous enhancement of FF-type pairing fluctuations. In this paper, we extend these previous studies \cite{Kawamura2020, Kawamura2020_JLTP} to the superfluid phase and derive the quantum kinetic equation, to clarify the properties and stabilities of the unconventional Fermi superfluid state associated with this FF-type pairing.
\par
This paper is organized as follows. In Sec. II, we extend the BCS theory to the non-equilibrium steady state, by employing the Nambu-Keldysh Green's function technique. We also give concrete experimental setup to realized our proposal, both in ultracold Fermi gas and electron systems. In Sec. III, using the same technique, we derive a quantum kinetic equation to evaluate the time evolution of the superfluid order parameter, under the initial condition that it slightly deviates from the mean-field value. We show our results in Sec. IV. We first show possible mean-field solutions for non-equilibrium superfluid steady states. We then assess their stability from the time evolution of the superfluid order parameter, to draw the phase diagram in Figs. \ref{fig2}(a) and \ref{fig2}(b). Throughout this paper, we set $\hbar=k_{\rm B}=1$, and the system volume $V$ is taken to be unity, for simplicity.
\par
%%%%%%%%%%%%%%%%%%%%%%%%%%%%%%%%%%%%%%%%%%%%%%%%%%%%%%%%%%%%%%%%%%%%%%%%%%%%
%%%%%%%%%%%%%%%%%%%%%%%%%%%%%%%%%%%%%%%%%%%%%%%%%%%%%%%%%%%%%%%%%%%%%%%%%%%%
\par
%%%%%%%%%%%%%%%%%%%%%%%%%%%%%%%%%%%%%%%%%%%%%%%%%%%%%%%%%%%%%%%%%%%%%%%%%%%%
\section{BCS theory of non-equilibrium superfluid steady state}
\par
\subsection{Model driven-dissipative non-equilibrium Fermi gas}
\label{subsec:model}
\par
The model driven-dissipative two-component Fermi gas shown in Fig. \ref{fig1}(a) is described by the Hamiltonian,
\begin{equation}
H= H_{\rm sys} + H_{\rm env} + H_{\rm mix},
\label{eq.1}
\end{equation}
where each term has the form, in the Nambu representation \cite{Schrieffer, Nambu1960},
\begin{align}
H_{\rm sys} &= 
-\int d\bm{r}
\Psi^\dagger(\bm{r}) {\nabla_{\bm r}^2 \over 2m}\tau_3 \Psi(\bm{r}) 
\notag\\
&\hspace{0.4cm}
-U\int d\bm{r} 
\Psi^\dagger({\bm r})\tau_+\Psi({\bm r})
\Psi^\dagger({\bm r})\tau_-\Psi({\bm r}),
\label{eq_Hsys}
\\[4pt]
H_{\rm env} &= \sum_{\alpha={\rm L,R}}
\int d{\bm R}
\Phi_{\alpha}^\dagger(\bm{R})
\left[
-{\nabla_{\bm R}^2 \over 2m} -\omega_\alpha
\right]\tau_3 \Phi_{\alpha}(\bm{R}),
\label{eq_Henv}
\\
H_{\rm mix} &=\sum_{\alpha={\rm L}, {\rm R}} \sum_{i=1}^{N_{\rm t}}
\left[
\Lambda_\alpha \Phi_\alpha^\dagger(\bm{R}^\alpha_i)
e^{i\tau_3 \mu_\alpha t}\tau_3 \Psi(\bm{r}^\alpha_i) + {\rm H.c.} 
\right].
\label{eq_Hmix}
\end{align}
In these Hamiltonians, 
\begin{align}
\Psi({\bm r}) &=
\left(
\begin{array}{c}
\psi_\uparrow({\bm r}) \\
\psi_\downarrow^\dagger({\bm r}) \\
\end{array}
\right),
\label{eq.2}\\[4pt]
\Phi_\alpha({\bm R}) &=
\left(
\begin{array}{c}
\phi_{\alpha,\uparrow}({\bm R})\\
\phi_{\alpha,\downarrow}^\dagger({\bm R})
\end{array}
\right),
\label{eq.3}	
\end{align}
represent the two-component Nambu fields describing fermions in the main system and the $\alpha={\rm L,R}$ reservoir, respectively (where $\psi_\sigma({\bm r})$ ($\phi_{\alpha,\sigma}({\bm R})$) is the annihilation operator of a fermion with pseudo-spin $\sigma=\uparrow,\downarrow$ and particle mass $m$ in the main system ($\alpha$-reservoir)). The corresponding Pauli matrices $\tau_i$ ($i=1,2,3$), as well as $\tau_\pm=[\tau_1\pm i\tau_2]/2$, act on the particle-hole space.
\par
$H_{\rm sys}$ in Eq. (\ref{eq_Hsys}) describes the main system in Fig. \ref{fig1}(a), consisting of a two-component Fermi gas with an $s$-wave pairing interaction $-U~(<0)$. Because $H_{\rm sys}$ involves the ultraviolet divergence, as usual in cold Fermi gas physics, we remove this singularity by measuring the interaction strength in terms of the $s$-wave scattering length $a_s$ \cite{Randeria1995}. It is related to the pairing interaction $-U$ as
\begin{equation}
{4\pi a_s \over m} = -{U  \over \displaystyle 1-U \sum_{\bm p } {1 \over 2\ep_{\bm p}}},
\label{eq_U}
\end{equation}
where $\varepsilon_{\bm p}={\bm p}^2/(2m)$ is the kinetic energy of a Fermi atom with an atomic $m$. In this paper, we only deal with the weak-coupling regime, by setting $(p_{\rm F}a_s)^{-1}=-1$. Here, $p_{\rm F} = \sqrt{2m\mu}$, where $\mu\equiv[\mu_{\rm R}+\mu_{\rm L}]/2~(>0)$ is the averaged chemical potential between the two reservoirs (where $\mu_\alpha$ is the chemical potential of the $\alpha$-reservoir in Fig. \ref{fig1}(a)).
\par
%%%%%%%%%%%%%%%%%%%%%%%%%%%%%%%%%%%%%%%%%%%%%%%%%%%%%%%%%%%%%%%%%%%%%%%%%%%%
\begin{figure}[tb]
\centering
\includegraphics[width=8cm]{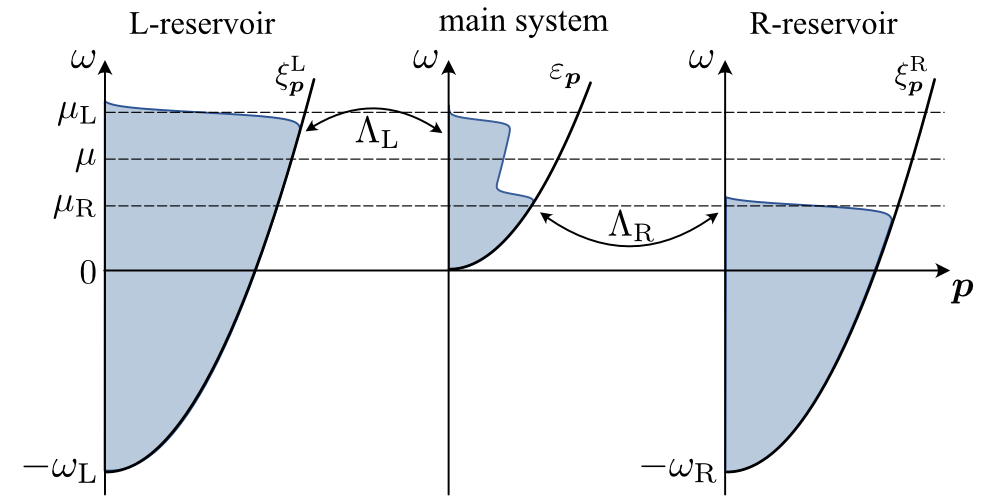}
\caption{
Schematic energy band structure of our model. We measure the energy from the bottom ($\ep_{\bm{p}=0}=0$) of the energy band in the main system. We set $\mu\ll\omega_\alpha$ so that the reservoirs are huge compared to the main system.
}
\label{fig3} 
\end{figure}
%%%%%%%%%%%%%%%%%%%%%%%%%%%%%%%%%%%%%%%%%%%%%%%%%%%%%%%%%%%%%%%%%%%%%%%%%%%%
\par
The two reservoirs ($\alpha={\rm L, R}$) are described by the Hamiltonian $H_{\rm env}$ in Eq. (\ref{eq_Henv}). As schematically shown in Fig. \ref{fig3}, $-\omega_{\alpha={\rm L}, {\rm R}}$ in Eq. (\ref{eq_Henv}) gives the bottom of the energy band in the $\alpha$-reservoir, when the energy is measured from the bottom ($\ep_{\bm{p}=0}=0$) of the energy band in the main system. We assume that both reservoirs are huge compared to the main system (which is satisfied by taking $\omega_\alpha$ to be sufficiently large compared to $\mu$), and are always in the thermal equilibrium state at the common environment temperature $T_{\rm env}$ \cite{noteT}. The particle occupation in each reservoir obeys the ordinary Fermi distribution function at $T_{\rm env}$,
\begin{equation}
f(\omega) = \frac{1}{e^{\omega/T_{\rm env}} +1}. 
\label{eq_barh_Fermi}
\end{equation}
\par
The coupling between the main system and the reservoirs is described by $H_{\rm mix}$ in Eq. (\ref{eq_Hmix}) with the coupling strength $\Lambda_{\alpha={\rm L},{\rm R}}$. For simplicity, we set $\Lambda_{\rm L}=\Lambda_{\rm R}\equiv\Lambda$ in what follows. The particle tunneling is assumed to occur between randomly distributing spatial positions $\bm{R}_i^\alpha$ in the $\alpha$-reservoir and ${\bm r}^\alpha_i$ in the main system ($i=1,\cdots N_{\rm t} \gg 1$). As we will see later (see Eq.~\eqref{eq:translational invariance} and the paragraphs nearby), the random tunneling points introduced here assures the  supply/decay of the particle from/to the reservoir to be uniform \cite{noteRandom}. This mimics the experimental situations that we propose below, where the injection of particles are also approximately uniform (to be discussed later). This phenomenological modeling parameters $\Lambda$ and $N_t$ would only appear in the expression of the linewidth $\gamma$ [see Eq.~\eqref{eq.pq}], which we interpret as a fitting parameter that would be determined experimentally \cite{Hanai2016,Hanai2017,Hanai2018}. 
\par
In Eq. \eqref{eq_Hmix}, the factor $\exp(i\tau_3 \mu_\alpha t)$ describes the situation that the energy band in the $\alpha$-reservoir is filled up to $\mu_\alpha$ (at $T_{\rm env}=0$), as schematically shown in Fig. \ref{fig3} \cite{Kawamura2020, Kawamura2020_JLTP}. By imposing the chemical-potential bias $\mu_{\rm L}=\mu+\delta\mu$ and $\mu_{\rm R}=\mu-\delta\mu$ ($\delta\mu>0$) between the two reservoirs, we realize the non-equilibrium steady state in the main system. In this paper, we fix the value of the average chemical potential $\mu$, and tune the non-equilibrium situation of the main system by adjusting the tunneling matrix element $\Lambda$, as well as the chemical-potential bias $\delta\mu$ \cite{noteP}.
\par
At the end of this subsection, we comment on this model and its feasibility in experiments. We expect the proposed setup can experimentally be realized, both in ultracold Fermi systems and electron systems. In the former systems, several groups have succeeded in splitting a cloud of a trapped Fermi gas into two \cite{Brantut2012, Krinner2015, Husmann2015, Krinner2017, Husmann_thesis, Kanasz-Nagy2016}. This system is known to be well described by two systems coupled via the tunneling between them \cite{Uchino2017, Yao2018, Sekino2020, Furutani2020}, similarly to our Hamiltonian (Eq.~\eqref{eq.1}). Thus, we strongly believe that our system composed of \textit{three} systems (the main system and the left and right reservoir) can be realized by extending this two-terminal setup into three, and applying a magnetic field to tune the strength of a pairing interaction associated with a Feshbach resonance only to the main system by using the techniques developed in \cite{Bauer2009, Yamazaki2010, Fu2013, Jagannathan2016, Arunkumar2018, Arunkumar2019}. In this setup, the tunneling process between the system and reservoirs would be overwhelmingly complicated: When a particle is injected from the reservoir to the system, they relax, decohere, and diffuse via inelastic collision processes. However, since this process seems to occur very quickly in the two-terminal configuration in cold atomic systems, the pumping can be regarded as being effectively uniform (which justifies our modeling with random tunneling points). This is supported by the verification of the Landauer formula \cite{Krinner2015} and the a.c. and d.c. Josephson effects \cite{Levy2007}, which uses the above assumption of fast dissipation over space. In Appendix C, we further argue that it is possible to achieve the parameter regime necessary to realize the unconventional states (i.e. $\gamma/\mu < 0.01$) within the current experimental techniques.
\par
In electron systems, a metal (that turns into a superconductor at low temperature) under a strong voltage bias ($eV\gg k_{\rm B} T$) is also a promising experimental setup to realize our proposed non-equilibrium FF-type superconducting state. In fact, a non-equilibrium quasiparticle distribution with the two-step structure analogous to the one depicted in Fig. \ref{fig1}(a) has been observed in voltage-biased mesoscopic wires \cite{Pothier1997, Poither1997_2, Anthore2003}, as well as in carbon nanotubes \cite{Chen2009}. The non-equilibrium quasiparticle distribution in a voltage-biased mesoscopic superconducting wire has also been investigated theoretically based on the quasiclassical theory, which has shown that a non-equilibrium distribution with the two-step structure is realized near the center of the wire, when the wire is sufficiently long compared to the superconducting coherence length \cite{Keizer2006, Vercruyssen2012}. Thus, we expect that the exotic superconducting state induced by the non-equilibrium quasiparticle distribution can also be observed in voltage-biased mesoscopic superconducting wires connected to normal-metal electrodes \cite{Boogaard2004, Li2011, Vercruyssen2012} or thin superconducting films sandwiched between electrodes \cite{Ohtomo2004, Reyren2007, Ueno2008}. In Appendix C, we provide further arguments on the feasibility of the realization of our proposal.
%%%%%%%%%%%%%%%%%%%%%%%%%%%%%%%%%%%%%%%%%%%%%%%%%%%%%%%%%%%%%%%%%%%%%%%%%%%%%%%%
\par
\begin{widetext}
\subsection{Non-equilibrium Nambu-Keldysh Green's function}
\par
To deal with the non-equilibrium superfluid state in the main system, we conveniently employ the $4\times 4$ matrix Nambu-Keldysh Green's function \cite{Rammer2007, Zagoskin2014}, given by
\begin{equation}
\hat{\m{G}}(1,2) =
\left(
\begin{array}{cc}
\m{G}^{\rm R}(1,2) & \m{G}^{\rm K}(1,2) \\
0 & \m{G}^{\rm A}(1,2)
\end{array}
\right)
=
\left(
\begin{array}{cc}
-i\Theta(t_1 -t_2)
\langle[\Psi(1)\diamondcomma \Psi^\dagger(2)]_+\rangle 
& -i\langle[\Psi(1)\diamondcomma \Psi^\dagger(2)]_-\rangle
\\
0 & i\Theta(t_2 -t_1)\langle[\Psi(1)\diamondcomma \Psi^\dagger(2) ]_+\rangle
\end{array}
\right),
\label{eq.4}
\end{equation}
\end{widetext}
where $\Theta(t)$ is the step function, and the abbreviated notations $1\equiv ({\bm r}_1, t_1)$ and $2\equiv ({\bm r}_2, t_2)$ are used. In Eq. (\ref{eq.4}), $\m{G}^{\rm R}$, $\m{G}^{\rm A}$, and $\m{G}^{\rm K}$ are the $2\times 2$ matrix retarded, advanced, and Keldysh Green's functions in the two-component Nambu space, respectively. In Eq. (\ref{eq.4}),
\begin{equation}
[\Psi(1)\diamondcomma \Psi^\dagger(2) ]_\pm=
\Psi(1)\diamond\Psi^\dagger(2)\pm\Psi^\dagger(2)\diamond\Psi(1),
\end{equation}
where ``$\diamond$" denotes the operation,
\begin{align}
&\Psi(1)\diamond\Psi^\dagger(2)=
\left(
\begin{array}{cc}
\psi_\up(1) \psi^\dagger_\up(2) & \psi_\up(1) \psi_\down(2) \\
\psi_\down^\dagger(1) \psi_\up^\dagger(2) & \psi_\down^\dagger(1) \psi_\down(2)
\end{array}
\right), \\[4pt]
& \Psi^\dagger(2) \diamond\Psi(1)=
\left(
\begin{array}{cc}
\psi^\dagger_\up(2) \psi_\up(1) & \psi_\down(2) \psi_\up(1) \\
\psi^\dagger_\up(2) \psi^\dagger_\down(1) & \psi_\down(2) \psi^\dagger_\down(1)
\end{array}
\right).
\label{eq.amplitude}
\end{align}
For later convenience, we also introduce the $4\times 4$ matrix lesser Green's function $\m{G}^<(1,2)$, which is related to $\m{G}^{\rm R,K,A}(1,2)$ as
\begin{align}
\m{G}^<(1,2)
&=
i\braket{\Psi^\dagger(2)\diamond\Psi(1)}
\notag\\
&= 
\frac{1}{2}\big[
\m{G}^{\rm K}(1,2)-\m{G}^{\rm R}(1,2) +\m{G}^{\rm A}(1,2)
\big].
\label{eq_GL}
\end{align}
We briefly note that the diagonal and off-diagonal components of $\m{G}^<$ are related to the particle density and the pair amplitude, respectively. 
\par
In the Nambu-Keldysh scheme, effects of the pairing interaction $-U$ and the system-reservoir couplings $\Lambda_{\alpha={\rm L,R}}$ can be summarized by the $4\times 4$ matrix self-energy correction,
\begin{equation}
\hat{\Sigma}(1,2) = 
\left(\begin{array}{cc}
\Sigma^{\rm R}(1,2) &\Sigma^{\rm K}(1,2)  \\[4pt]
0 & \Sigma^{\rm A}(1,2)
\end{array} \right) =
\hat{\Sigma}_{\rm int}(1,2) + \hat{\Sigma}_{\rm env}(1,2), 
\label{eq_self}
\end{equation}
which appears in the non-equilibrium Nambu-Keldysh Dyson equation \cite{Rammer2007, Zagoskin2014},
\begin{equation}
\hat{\m{G}}(1,2) =\hat{\m{G}}_0(1,2) +\big[\hat{\m{G}}_0 \circ \hat{\Sigma} \circ \hat{\m{G}} \big](1,2).
\label{eq_Dyson}
\end{equation}
Here,
\begin{align}
\scalebox{0.92}{$\displaystyle
\big[A \circ B\big](1,2)  $}
&= 
\scalebox{0.92}{$\displaystyle
\int d\bm{r}_3 \int_{-\infty}^{\infty}dt_3 
A(\bm{r}_1,t_1,\bm{r}_3,t_3) 
B(\bm{r}_3,t_3,\bm{r}_2,t_2) $}
\notag\\
&= \int d3 A(1,3) B(3,2),
\label{eq.circ}
\end{align}
and 
\begin{align}
\hat{\m{G}}_0(1,2)
&=
\sum_{\bm p}\int{d\omega \over 2\pi}
e^{i{\bm p}\cdot({\bm r}_1-{\bm r}_2)-i\omega(t_1-t_2)}
\hat{\m{G}}_0({\bm p},\omega)
\nonumber
\\
&=
\sum_{\bm p}\int{d\omega \over 2\pi}
e^{i{\bm p}\cdot({\bm r}_1-{\bm r}_2)-i\omega(t_1-t_2)}
\nonumber
\\
&\hspace{0.4cm}\times
\left(\begin{array}{cc}
\frac{1}{\omega_+ -\ep_{\bm{p}}\tau_3} & -2\pi i \delta(\omega -\ep_{\bm{p}}) \big[1 -2f_{\rm ini}(\omega) \big]\\
0&\frac{1}{\omega_- -\ep_{\bm{p}}\tau_3}
\end{array}\right)
\label{eq.5}
\end{align}
is the bare Green's function in the {\it initial} thermal equilibrium state at $t=-\infty$, where the system-reservoir couplings $\Lambda_{\alpha={\rm L,R}}$, as well as the pairing interaction $-U$, were absent. (Note that $\hat{\m{G}}_0(1,2)$ only depends on the relative coordinate as $\hat{\m{G}}_0(1,2)=\hat{\m{G}}_0(1-2)$.) In Eq. (\ref{eq.5}), $\omega_\pm=\omega\pm i\delta$ (where $\delta$ is an infinitesimally small positive number), and $f_{\rm ini}(\omega)=1/[e^{\omega/T_{\rm ini}} +1]$ is the Fermi distribution function with $T_{\rm ini}$ being the initial temperature of the main system at $t=-\infty$. Although the Dyson equation (\ref{eq_Dyson}) looks like depending on the initial state through $\hat{\m{G}}_0(1,2)$, we will soon find that the dressed Green's function $\hat{\m{G}}(1,2)$ in the final non-equilibrium steady state, which we are interested in, actually loses the initial memory \cite{Kawamura2020,  Hanai2016, Hanai2017, Hanai2018, Kawamura2020_JLTP, Szymanska2006,Szymanska2007, Yamaguchi2012,Yamaguchi2015}. 
\par
\begin{figure}[tb]
\centering
\includegraphics[width=7.8cm]{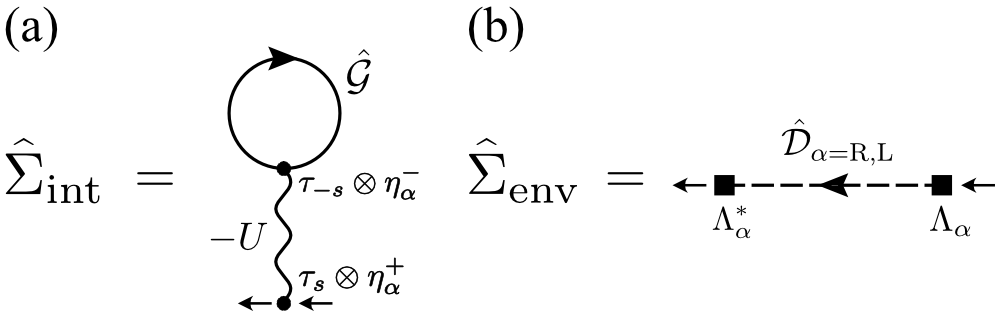}
\caption{Self-energy corrections. (a) $\hat{\Sigma}_{\rm int}$ describes effects of the pairing interaction $-U$ in the mean-field BCS approximation. The solid line is the dressed Nambu-Keldysh Green's function $\hat{\m{G}}$ in the main system. The wavy line is the pairing interaction $-U$, which is accompanied by the vertices $\tau_{\pm s} \otimes \eta^\pm_\alpha$ at both ends (where $s=\pm$), acting on the Nambu $\otimes$ Keldysh space. (b) $\hat{\Sigma}_{\rm env}$ describes effects of the system-reservoir couplings $\Lambda_{\alpha={\rm L, R}}$ in the second-order Born approximation. The dashed line is the Green's function $\hat{\m{D}}_{\alpha={\rm L, R}}$ in the $\alpha$-reservoir. The solid square represents the tunneling matrix $\Lambda_{\alpha={\rm L,R}}$ between the system and the $\alpha$-reservoir.}
\label{fig_diagram} 
\end{figure}

In Eq. (\ref{eq_self}), $\hat{\Sigma}_{\rm int}$ and $\hat{\Sigma}_{\rm env}$  describe effects of the pairing interaction $-U$ and the system-reservoir couplings $\Lambda_{\alpha={\rm L,R}}~(=\Lambda)$, respectively. In the mean-field BCS approximation \cite{Hanai2016, Hanai2017, Hanai2018}, the former is diagrammatically drawn as Fig. \ref{fig_diagram}(a), which gives
\begin{widetext}
\begin{align}
& \hat{\Sigma}_{\rm int}(1,2)  = iU \sum_{s=\pm} \sum_{\alpha= 1,2}  \big( \tau_s \otimes \eta^+_\alpha \big) {\rm Tr}_{\rm N}{\rm Tr}_{\rm K}\big[ \big(\tau_{-s} \otimes \eta^-_{\alpha}\big) \hat{\m{G}}(1,2) \big] \delta(1-2)
\notag\\[4pt]
&= 
\frac{iU}{2}\sum_{s=\pm}
\left(\begin{array}{cc}
\tau_s {\rm Tr}_{\rm N}\big[ \tau_{-s} \m{G}^{\rm K}(1, 2) \big] &
\tau_s {\rm Tr}_{\rm N}\big[ \tau_{-s} \m{G}^{\rm R}(1, 2) + \tau_{-s} \m{G}^{\rm A}(1, 2)  \big] \\[4pt]
\tau_s {\rm Tr}_{\rm N}\big[ \tau_{-s} \m{G}^{\rm R}(1, 2) + \tau_{-s} \m{G}^{\rm A}(1, 2) \big] &
\tau_s {\rm Tr}_{\rm N}\big[ \tau_{-s} \m{G}^{\rm K}(1, 2) \big]
\end{array}\right)\delta(1-2).
\label{NK_self_int0}
\end{align}
Here, 
\begin{equation}
\eta_\alpha^+ = \frac{1}{\sqrt{2}} \sigma_{2-\alpha}, {\hskip 0.5cm} \eta_\alpha^- = \frac{1}{\sqrt{2}} \sigma_{\alpha-1}
\end{equation}
are vertex matrices \cite{Kawamura2020}, where $\sigma_{i=1,2,3}$ are the Pauli matrices acting on the Keldysh space. ${\rm Tr}_{\rm N}$ and ${\rm Tr}_{\rm K}$ stand for taking the trace over the Nambu and the Keldysh space, respectively.
\par
We introduce the superfluid order parameter $\Delta(\bm{r}_1, t_1)$, which is related to the off-diagonal component of $\m{G}^{\rm K}$ and $\m{G}^<$ as
\begin{equation}
\Delta(\bm{r}_1, t_1) 
\equiv U \braket{\psi_\down(1) \psi_\up (1)} = -\frac{iU}{2} \m{G}^{\rm K}(1,1)_{12}
=-iU\m{G}^{<}(1,1)_{12}.
\label{def_OP}
\end{equation}
Then, Eq. (\ref{NK_self_int0}) can be simply written as
\begin{align}
\hat{\Sigma}_{\rm int}(1,2)
=&
\left(\begin{array}{cc}
- \Delta(1) \tau_+ -\Delta^*(1) \tau_- & 0 \\[4pt]
0 & - \Delta(1) \tau_+ -\Delta^*(1) \tau_- 
\end{array}\right)
\delta(1-2). 
\label{NK_self_int1}
\end{align}
We note that the off-diagonal components of $\hat{\Sigma}_{\rm int}(1,2)$ identically vanish, because $\m{G}^{\rm R(A)}(1,1)_{12}=\m{G}^{\rm R(A)}(1,1)_{21}=0$. 
\par
For the self-energy correction $\hat{\Sigma}_{\rm env}$ in Eq. (\ref{eq_self}), we take into account the system-reservoir couplings within the second-order Born approximation, as diagrammatically shown in Fig. \ref{fig_diagram}(b). Evaluation of this diagram gives

\begin{eqnarray}
\hat{\Sigma}_{\rm env}({\bm p},{\bm p}',t_1,t_2)
&=&
\int d{\bm r}_1
\int d{\bm r}_2
\hat{\Sigma}_{\rm env}(1,2) 
e^{-i({\bm p}\cdot{\bm r}_1+{\bm p}'\cdot{\bm r}_2)}
\nonumber\\ 
&=& |\Lambda|^2 
\sum_{\alpha={\rm L,R}}
\sum_{i,j}^{N_{\rm t}}
\hat{\m{D}}_{\alpha}(\bm{R}_i^\alpha- \bm{R}_j^\alpha, t_1-t_2 ) 
e^{-i\tau_3\mu_\alpha(t_1-t_2)}
e^{-i({\bm p}\cdot{\bm r}_i+{\bm p}'\cdot{\bm r}_j)},
\label{NK_self_env1}
\end{eqnarray}
where 
\begin{align}
\hat{\m{D}}_{\alpha}(\bm{R}_i^\alpha-\bm{R}_j^\alpha, t_1-t_2 )
&=
\sum_{\bm q}\int{d\omega \over 2\pi}
e^{i{\bm q}\cdot({\bm R}_i^\alpha-{\bm R}_j^\alpha)-i\omega(t_1-t_2)}
\hat{\m{D}}_0({\bm q},\omega)
\nonumber
\\
&=\sum_{\bm q}\int{d\omega \over 2\pi}
e^{i{\bm q}\cdot({\bm R}_i^\alpha-{\bm R}_j^\alpha)-i\omega(t_1-t_2)}
\left( 
\begin{array}{cc}
\frac{1}{\omega_+ -\xi^\alpha_{\bm{q}}\tau_3} & 
-2\pi i \delta(\omega -\xi^\alpha_{\bm{q}}) \tanh\left( \frac{\omega}{2T_{\rm env}} \right)  \\
0 & \frac{1}{\omega_- -\xi^\alpha_{\bm{q}}\tau_3}
\end{array}
\right)
\label{eq.6}
\end{align}
\end{widetext}
is the non-interacting Nambu-Keldysh Green's function in the $\alpha$-reservoir, with $\xi_{\bm{q}}^\alpha=\ep_{\bm{q}} -\omega_\alpha$ being the kinetic energy measured from the bottom energy $-\omega_\alpha$ of the band in the $\alpha$-reservoir. (Note that the reservoirs are assumed to be in thermal equilibrium.) When one takes the spatial averages over the randomly distributing tunneling positions $\bm{R}_i^\alpha$ and $\bm{r}_i^\alpha$ in Eq. \eqref{NK_self_env1}, the resulting self-energy $\braket{{\hat \Sigma}_{\rm env}({\bm p},{\bm p}',t_1,t_2)}_{\rm av}$ recovers its the translational invariance as \cite{Hanai2016,Hanai2017,Hanai2018,Kawamura2020}
\begin{equation}
\label{eq:translational invariance}
\braket{{\hat \Sigma}_{\rm env}({\bm p},{\bm p}',t_1,t_2)}_{\rm av}=\hat{\Sigma}_{\rm env}({\bm p},t_1,t_2)\delta_{{\bm p},{\bm p}'}, 
\end{equation}
where
\begin{equation}
\scalebox{0.95}{$\displaystyle
\hat{\Sigma}_{\rm env}({\bm p},t_1,t_2)=N_t|\Lambda|^2 \sum_{{\bm q},\alpha={\rm L,R}} \hat{\m{D}}_{\alpha}({\bm q}, t_1-t_2 ) e^{-i\mu_\alpha(t_1-t_2)\tau_3}. $}
\label{NK_self_env2}
\end{equation}
Carrying out the Fourier transformation with respect to the relative time $t_1 -t_2$, we have
\begin{equation}
\hat{\Sigma}_{\rm env}(\bm{p},\omega) = N_{\rm t} |\Lambda|^2 
\sum_{{\bm q},\alpha={\rm L,R}}
\hat{\m{D}}_\alpha\big(\bm{q}, \omega-\mu_\alpha\tau_3 \big).
\label{NK_self_env3}
\end{equation}
\par
%%%%%%%%%%%%%%%%%%%%%%%%%%%%%%%%%%%%%%%%%%%%%%%%%%%%%%%%%%%%%%%%%%%%%%%%%%%%%%%%
%%%%%%%%%%%%%%%%%%%%%%%%%%%%%%%%%%%%%%%%%%%%%%%%%%%%%%%%%%%%%%%%%%%%%%%%%%%%%%%%
\par
For simplicity, we employ the so-called wide-band limit approximation \cite{Stefanucci2013}, that is, we assume white reservoirs with the constant density of states $\rho_\alpha(\omega) \equiv \rho$. This approximation is justified when the reservoirs are so huge that the energy dependence of the reservoir density of states around the Fermi level can be ignored \cite{Kawamura2020}, which is just the situation we are considering ($\mu\ll\omega_\alpha$). Then, replacing $\bm{q}$ summation in Eq. \eqref{NK_self_env3} by the $\xi^\alpha$ integration, one has 
\begin{align}
&\hat{\Sigma}_{\rm env}(\bm{p},\omega) 
\notag\\
&=
\scalebox{0.93}{$\displaystyle
\left(\begin{array}{cc}
-2i\gamma \tau_0 & -2i\gamma\left[\tanh\left( \frac{\omega-\tau_3\mu_{\rm L}}{2T_{\rm env}} \right) +\tanh\left( \frac{\omega-\tau_3\mu_{\rm R}}{2T_{\rm env}} \right)\right]\tau_0\\
0 &2i\gamma \tau_0
\end{array} \right).  $}
\label{NK_self_env4}
\end{align}
Here, 
\begin{equation}
\gamma=\pi N_{\rm t}\rho|\Lambda|^2
\label{eq.pq}
\end{equation}
is the quasi-particle damping rate, and $\tau_0$ is the $2\times 2$ unit matrix acting on the particle-hole Nambu space.
\par
%%%%%%%%%%%%%%%%%%%%%%%%%%%%%%%%%%%%%%%%%%%%%%%%%%%%%%%%%%%%%%%%%%%%%%%%%%%%%%%%
\par
\subsection{Extension of BCS Theory to non-equilibrium steady state}
\par
In this paper, we explore stable non-equilibrium superfluid steady states, having the following type of the order parameter:
\begin{equation}
\Delta(\bm{r},t)=\Delta_0e^{i \bm{Q}\cdot \bm{r}}e^{-2i\mu t}.
\label{eq_OP}
\end{equation}
Without loss of generality, we take $\Delta_0>0$. When ${\bm Q}=0$, Eq. (\ref{eq_OP}) describes the BCS-type uniform superfluid, which has been discussed in exciton(-polariton) systems \cite{Szymanska2006, Hanai2016, Hanai2017, Hanai2018, Hanai2019, Yamaguchi2012, Yamaguchi2015}. When ${\bm Q}\ne 0$, Eq. (\ref{eq_OP}) has the same form as the order parameter in the Fulde-Ferrell (FF) superfluid state, discussed in superconductivity under an external magnetic field \cite{Fulde1964, Larkin1964, Takada1969, Shimahara1994, Matsuda2007}, as well as in a spin-polarized Fermi gas \cite{Hu2006, Parish2007, Liao2010, Chevy2010, Kinnunen2018, Strinati2018}. Although the Larkin-Ovchinnikov type solution \cite{Larkin1964}, $\Delta(\bm{r},t)=\Delta_0 \cos(\bm{Q}\cdot \bm{r}) e^{-2i\mu t}$, is also conceivable in our model, leaving this possibility as our future study, we only deal with the FF-type solution in this paper. Regarding this, we emphasize that the main system has {\it no} spin imbalance.
\par
To treat the superfluid order parameter $\Delta(\bm{r}, t)$ in Eq. (\ref{eq_OP}), it is convenient to formally remove time and spatial dependence from it, which is achieved by employing the following gauge transformation \cite{Rammer2007}:
\begin{align}
\hat{\tilde{\m{G}}}(1,2)&=
\left(
\begin{array}{cc}
{\tilde {\m G}}^{\rm R}(1,2) &
{\tilde {\m G}}^{\rm K}(1,2) \\
0 &
{\tilde {\m G}}^{\rm A}(1,2) 
\end{array}
\right)
\notag\\[4pt]
&\equiv
 e^{-i\chi(1) \tau_3\otimes \sigma_0} \hat{\m{G}}(1, 2) e^{i\chi(2) \tau_3\otimes\sigma_0},
\label{eq.gaugeA}
\\[8pt]
\hat{\tilde{\Sigma}}(1, 2)&=
\left(
\begin{array}{cc}
{\tilde \Sigma}^{\rm R}(1,2) &
{\tilde \Sigma}^{\rm K}(1,2) \\
0 &
{\tilde \Sigma}^{\rm A}(1,2) 
\end{array}
\right)
\notag\\[4pt]
&\equiv
e^{-i\chi(1) \tau_3\otimes \sigma_0}\hat{\Sigma}(1,2)e^{i\chi(2) \tau_3\otimes\sigma_0},
\label{eq.gaugeB}
\end{align}
where 
\begin{equation}
\chi(\bm{r}, t)= {1 \over 2}\bm{Q} \cdot \bm{r} -\mu t,
\label{eq.gaugeC}
\end{equation}
and $\sigma_0$ is the $2\times 2$ unit matrix acting on the Keldysh space. The Nambu-Keldysh Dyson equation (\ref{eq_Dyson}) after this manipulation is given by
\begin{equation}
\hat{\tilde{\m{G}}}(1,2) =\hat{\tilde{\m{G}}}_0(1,2) +\big[\hat{\tilde{\m{G}}}_0 \circ \hat{\tilde{\Sigma}} \circ \hat{\tilde{\m{G}}} \big](1,2). 
\label{eq_Dyson_tilde}
\end{equation}
Here,
\begin{equation}
\hat{\tilde{\m{G}}}_0(1,2) =\hat{\tilde{\m{G}}}_0(1-2) =
e^{i[\mu(t_1-t_2)-{\bm{Q} \over 2} \cdot (\bm{r}_1 -\bm{r}_2)]\tau_3}
\hat{\m{G}}_0(1-2).
\end{equation}
is the gauge-transformed bare Green's function. In the energy and momentum space, Eq. (\ref{eq_Dyson_tilde}) has the form,
\begin{align}
\hat{\tilde{\m{G}}}(\bm{p}, \omega)
&=
\left(
\begin{array}{cc}
\tilde{\m{G}}^{\rm R}({\bm p},\omega) & 
\tilde{\m{G}}^{\rm K}({\bm p},\omega)\\
0& \tilde{\m{G}}^{\rm A}({\bm p},\omega)
\end{array}
\right)
\notag\\[4pt]
&=
\hat{\tilde{\m{G}}}_0(\bm{p}, \omega) +\hat{\tilde{\m{G}}}_0(\bm{p}, \omega) \hat{\tilde{\Sigma}}(\bm{p}, \omega) \hat{\tilde{\m{G}}}(\bm{p}, \omega),
\label{eq_NESS_Dyson}
\end{align}
where 
\begin{widetext}
\begin{align}
\hat{\tilde{\m{G}}}_0(\bm{p}, \omega)&=
\int d{\bm r}\int dt e^{i[\omega t-{\bm p}\cdot{\bm r}]} 
\hat{\tilde{\m{G}}}_0({\bm r},t)
=
\hat{\m{G}}_0(\bm{p} +{\bm{Q} \over 2}\tau_3, \omega +\mu\tau_3) 
\notag\\[8pt]
&=
\left( \begin{array}{cc}
\frac{1}{\omega_+ -\xi_{\bm{p} +(\bm{Q}/2)\tau_3}\tau_3} & 
-2\pi i \delta(\omega -\xi_{\bm{p} +(\bm{Q}/2)\tau_3}\tau_3)
\big[ 1- 2f_{\rm ini}(\omega +\mu\tau_3)\big]  \\
0 & \frac{1}{\omega_- -\xi_{\bm{p} +(\bm{Q}/2)\tau_3}\tau_3}
\end{array}\right), 
\label{tilde_G0}
\end{align}
\end{widetext}
with $\xi_{\bm{p}}=\ep_{\bm{p}}-\mu$. The interaction component $\hat{\tilde{\Sigma}}_{\rm int}$ of the self-energy $\hat{\tilde{\Sigma}}=\hat{\tilde{\Sigma}}_{\rm int}+\hat{\tilde{\Sigma}}_{\rm env}$ in Eq. (\ref{eq_Dyson_tilde}) is obtained from Eq. \eqref{NK_self_int1} as
\begin{equation}
\hat{\tilde{\Sigma}}_{\rm int}(1,2)=\hat{\tilde{\Sigma}}_{\rm int}(1-2) =
\left(\begin{array}{cc}
-\Delta_0 \tau_1 & 0 \\
0 & -\Delta_0 \tau_1
\end{array}\right) \delta(1-2). 
\end{equation}
In the energy and momentum space, this self-energy has the form,
\begin{equation}
\hat{\tilde{\Sigma}}_{\rm int}(\bm{p}, \omega) =
\left(\begin{array}{cc}
-\Delta_0 \tau_1 & 0 \\
0 & -\Delta_0 \tau_1
\end{array}\right). 
\label{tilde_self_int}
\end{equation}
In the same manner, the self-energy $\hat{\tilde{\Sigma}}_{\rm env}(1,2)$ coming from the system-reservoir couplings can also be obtained from Eq. \eqref{NK_self_env4}. Thus, it has the form, in the energy and momentum space,
\begin{align}
&
\hat{\tilde{\Sigma}}_{\rm env}(\bm{p},\omega) = \hat{\Sigma}_{\rm env}(\bm{p}+{\bm{Q} \over 2}\tau_3, \omega +\mu\tau_3)
\notag\\
&=
\left(\begin{array}{cc}
-2i\gamma \tau_0 & -2i\gamma\left[\tanh\left( \frac{\omega-\delta\mu}{2T_{\rm env}} \right) +\tanh\left( \frac{\omega+\delta\mu}{2T_{\rm env}} \right)\right]\tau_0\\
0 &2i\gamma \tau_0
\end{array} \right).
\label{self_tilde_env}
\end{align}
The self-energy $\hat{\tilde{\Sigma}}(\bm{p}, \omega)$ involved in the Dyson equation (\ref{eq_NESS_Dyson}) is then given by the sum of Eqs. (\ref{tilde_self_int}) and (\ref{self_tilde_env}).
\par
Solving the Dyson equation (\ref{eq_NESS_Dyson}), we obtain
\begin{align}
\tilde{\m{G}}^{\rm R}(\bm{p}, \omega) &=
\sum_{\eta=\pm}
\frac{1}{\omega + 2i\gamma -E^{0,\eta}_{\bm{p},\bm{Q}}} \Xi
^{0,\eta}_{\bm{p},\bm{Q}}, 
\label{eq_NESS_GR}
\\
\tilde{\m{G}}^{\rm A}(\bm{p}, \omega) &=
\sum_{\eta=\pm}
\frac{1}{\omega - 2i\gamma -E^{0,\eta}_{\bm{p},\bm{Q}}} \Xi^{0,\eta}_{\bm{p},\bm{Q}}, 
\label{eq_NESS_GA}
\\
\tilde{\m{G}}^{\rm K}(\bm{p}, \omega) &=
\sum_{\eta=\pm}
\frac{4i\gamma[1-2F(\omega)]}{[\omega -\eta E^{0,\eta}_{\bm{p},\bm{Q}}]^2 +4\gamma^2} \Xi^{0,\eta}_{\bm{p},\bm{Q}},
\label{eq_NESS_GK}
\end{align}
where
\begin{align}
&
E^{0,\pm}_{\bm{p},\bm{Q}} = \sqrt{\bigl( \xi^{\rm (s)}_{\bm{p},\bm{Q}}\bigr)^2 + \Delta_0^2} \pm \xi^{\rm (a)}_{\bm{p},\bm{Q}} \equiv E^0_{\bm{p}, \bm{Q}} \pm \xi^{\rm (a)}_{\bm{p},\bm{Q}}, 
\label{eq_qp_exc}
\\[4pt]
&
\Xi^{0,\pm}_{\bm{p},\bm{Q}} = {1 \over 2}
\left[
\tau_0
\pm{\xi_{{\bm p},{\bm Q}}^{\rm (s)} \over E^0_{\bm{p}, \bm{Q}}}\tau_3
\mp{\Delta_0 \over E^0_{\bm{p}, \bm{Q}}}\tau_1
\right],
\end{align}
with $\xi^{\rm (s)}_{\bm{p},\bm{Q}}=[ \xi_{\bm{p}+\bm{Q}/2} + \xi_{-\bm{p}+\bm{Q}/2} ]/2$, and $\xi^{\rm (a)}_{\bm{p},\bm{Q}}= [\xi_{\bm{p}+\bm{Q}/2} - \xi_{-\bm{p}+\bm{Q}/2} ]/2$. In Eq. (\ref{eq_NESS_GK}), 
\begin{equation}
F(\omega)={1 \over 2}[f(\omega+\delta\mu)+f(\omega-\delta\mu)]
\label{eq.F}
\end{equation}
works as the non-equilibrium distribution function in the main system (although  $T_{\rm env}$ is used in $f(\omega)$, see Eq. (\ref{eq_barh_Fermi})). When we set $\omega=\xi_{\bm p}$, the resulting momentum distribution $F(\xi_{\bm p})$ has two Fermi-surface-like edges at $p_{\rm F1}=\sqrt{2m\mu_{\rm R}}$ and $p_{\rm F2}=\sqrt{2m\mu_{\rm L}}$, as schematically drawn in Fig. \ref{fig1}(a).
\par
The superfluid order parameter $\Delta({\bm r},t)$ in Eq. (\ref{eq_OP}) is self-consistently determined from Eq. (\ref{def_OP}). To evaluate this equation, we note that the gauge-transformed lesser Green's function $\tilde{\m{G}}^{<}(\bm{p}, \omega)$ in the energy and momentum space is obtained from Eqs. (\ref{eq_GL}) and (\ref{eq_NESS_GR})-(\ref{eq_NESS_GK}) as
\begin{align}
\tilde{\m{G}}^{<}(\bm{p}, \omega) &= \frac{1}{2}
\big[
\tilde{\m{G}}^{\rm K}(\bm{p}, \omega) -\tilde{\m{G}}^{\rm R}(\bm{p}, \omega) +\tilde{\m{G}}^{\rm A}(\bm{p}, \omega)
\big]
\notag\\ 
&=
\sum_{\eta=\pm}\frac{4i\gamma F(\omega)}{[\omega - \eta E^{0,\eta}_{\bm{p},\bm{Q}}]^2 +4\gamma^2} \Xi^{0,\eta}_{\bm{p},\bm{Q}}. 
\label{eq_NESS_GL}
\end{align}
Using Eq. (\ref{eq_NESS_GL}) in evaluating Eq. (\ref{def_OP}), we obtain the gap equation,
\begin{align}
1&=U\sum_{\bm{p}} \int_{-\infty}^\infty \frac{d\omega}{2\pi} 
\notag\\
&\hspace{0.6cm} \times
\frac{4\gamma \big[\omega -\xi^{\rm (a)}_{\bm{p},\bm{Q}} \big]\big[1-2F(\omega)\big]}{\bigl[(\omega- E^{0+}_{\bm{p},\bm{Q}})^2 + 4\gamma^2\bigr]\bigl[(\omega+E^{0-}_{\bm{p},\bm{Q}})^2 + 4\gamma^2\bigr]}. 
\label{eq_NESS_gap}
\end{align}
To quickly see the relation to the ordinary BCS gap equation, we take the thermal equilibrium and uniform limit, by setting $(\gamma,\delta\mu,{\bm Q})\to(+0,0,0)$. Then, Eq. (\ref{eq_NESS_gap}) is reduced to 
\begin{equation}
1= U\sum_{\bm{p}} \frac{1}{2E_{\bm p}}
\tanh
\left( 
\frac{E_{\bm p}}{2T_{\rm env}} 
\right),
\label{eq.BCSgap}
\end{equation}
where $E_{\bm p}=E_{{\bm p},{\bm Q}=0}^0=\sqrt{\xi_{\bm p}^2+\Delta_0^2}$. Equation (\ref{eq.BCSgap}) is just the same form as the ordinary BCS gap equation \cite{Schrieffer}, when one interprets $\mu$ and $T_{\rm env}$ as the Fermi chemical potential and the temperature in the main system, respectively. Thus, Eq. (\ref{eq_NESS_gap}) may be viewed as a non-equilibrium extension of the BCS gap equation. 
\par
Because $\Delta({\bm r},t)$ in Eq. (\ref{eq_OP}) involves two parameters $\Delta_0$ and ${\bm Q}$, we actually need one more equation to completely fix these parameters \cite{note}. For this purpose, we impose the vanishing condition for the net current ${\bm J}_{\rm net}$ in the main system. In the Nambu-Keldysh formalism, it is given by
\begin{equation}
\bm{J}_{\rm net} = \sum_{\sigma=\up,\down} \sum_{\bm{p}}\bigl[\bm{p}+\bm{Q}/2\bigr]n_{\bm{p},\sigma}=0,
\label{eq_NESS_current}
\end{equation}
where the Fermi momentum distribution $n_{\bm{p},\sigma}$ in the pseudo-spin $\sigma$ component is related to the diagonal component of the lesser Green's function $\tilde{\m{G}}^<(\bm{p}, \omega)$ in Eq. (\ref{eq_NESS_GL}) as
\begin{align}
n_{\bm{p},\up}&= -i\int_{-\infty}^{\infty} \frac{d\omega}{2\pi} \tilde{\m{G}}^<(\bm{p}, \omega)_{11}, \\
n_{\bm{p},\down}&= 1-i\int_{-\infty}^{\infty} \frac{d\omega}{2\pi} \tilde{\m{G}}^<(-\bm{p}, \omega)_{22}.
\end{align}
We solve the gap equation (\ref{eq_NESS_gap}) under the condition in Eq. (\ref{eq_NESS_current}), to self-consistently determine $(\Delta_0,{\bm Q})$ for a given set $(\gamma, \delta\mu, T_{\rm env})$ of environment parameters.
\par
We note that the net current $\bm{J}_{\rm net}$ is the current flowing \textit{inside} the main system, but {\it not} to be confused with the current flowing from the reservoir to the main system (the latter is always nonzero unless we consider the chemical equilibrium case $\mu_{\rm L}=\mu_{\rm R}$). Although we assume here that the former is zero, strictly speaking, there is no a priori way of determining it since it ultimately depends on the choice of a boundary condition \cite{noteBloch}. Hence, one may also consider more general cases with ${\bm J}_{\rm net}\ne 0$. Even in such a case, however, the essential physics remains unchanged: The non-equilibrium FF-type superfluid state with a non-zero net current can be realized as a stable non-equilibrium steady state (unless the current exceeds the Landau velocity, which would destroy all superfluid states). Thus, we choose the simplest case of the vanishing-current boundary condition with Eq. (\ref{eq_NESS_current}) for the sake of brevity.
\par
We also note that the normal state ($\Delta_0=\bm{Q}=0$) always satisfies the gap equation (\ref{eq_NESS_gap}) and the vanishing current condition in Eq. (\ref{eq_NESS_current}). However, as pointed out in our previous work \cite{Kawamura2020, Kawamura2020_JLTP}, the normal state becomes unstable, when the particle-particle scattering vertex $\chi(\bm{Q},\nu)$ develops a pole at $\nu=2\mu$. Here, ${\bm Q}$ and $\nu$ are the center-of-mass momentum and the total energy of two particles participating in the Cooper channel, respectively. Within the random phase approximation in terms of $-U$, one has \cite{Kawamura2020, Kawamura2020_JLTP}, in NESS, 
\begin{equation}
\chi(\bm{Q}, \nu=2\mu)=
{-U \over \displaystyle 1-{U \over 4\pi} \sum_{\eta, \zeta =\pm} 
{1 \over \xi^{\rm s}_{\bm{p},\bm{Q}}}{\rm Tan}^{-1}\left(\frac{\xi_{\bm{p},\bm{Q}}^{\eta,\zeta}}{2\gamma} \right)},
\label{eq.Thouless}
\end{equation}
where $\xi_{{\bm p},{\bm Q}}^{\eta,\zeta}=\xi_{\bm{p} +\eta \bm{Q}/2} + \zeta \delta\mu$. We determine the region where the normal state is stable on the phase diagram from the condition that $\chi(\bm{Q}, \nu=2\mu)<0$ for any $\bm{Q}$, that is, $\chi(\bm{Q},\nu)$ has no pole at $\nu=2\mu$.
\par
%%%%%%%%%%%%%%%%%%%%%%%%%%%%%%%%%%%%%%%%%%%%%%%%%%%%%%%%%%%%%%%%%%%%%%%%%%%%%%%%
\par
\section{Quantum kinetic theory to assess the stability of non-equilibrium Fermi superfluids}
\par
\subsection{Quantum kinetic equation}
\par
So far, we have obtained the self-consistent equations that the steady state solutions satisfy. In order to make sure that the obtained solutions are physical, we study the stability of such solutions. In this section, we explain how one can check such stability in a nonequilibrium setup.
\par
The stability of the obtained solution can be examined by tracing the time evolution of the superfluid order parameter under the initial condition that it is slightly deviated from the mean-field value at $t=0$. This is done here by deriving a quantum kinetic equation (QKE). We can then determine the stability of the steady-state solution by observing whether such deviation decays or grows over time.
\par
The central quantity of interest is the Wigner-transformed Nambu lesser Green's function \cite{Rammer2007,Zagoskin2014}, given by
\begin{equation}
\tilde{\m{G}}^<(\bm{p},\omega,\bm{r},t)
=\int d\bm{r}_{\rm r} \int dt_{\rm r} 
e^{i(\omega t_{\rm r}-\bm{p}\cdot \bm{r}_{\rm r})} 
\tilde{\m{G}}^<({\bm r}_1,t_1,{\bm r}_2,t_2).
\label{eq_Wigner_trans1}
\end{equation}
Here, $\bm{r}_{\rm r}=\bm{r}_1-\bm{r}_2$ and $\bm{r}=(\bm{r}_1+\bm{r}_2)/2$ are, respectively, the relative coordinate and the center-of-mass coordinate. For time variables, we have also introduced $t_{\rm r}=t_1 -t_2$ and $t=(t_1 + t_2)/2$ in Eq. (\ref{eq_Wigner_trans1}). It is useful to study such quantity since the off-diagonal component of this quantity is directly related to the (gauge-transformed) superfluid order parameter ${\bar \Delta}(\bm{r},t)$. This can be readily seen by summing up $\tilde{\m{G}}^<(\bm{p},\omega,\bm{r},t)$ over momentum $\bm p$ and frequency $\omega$:
\begin{align}
-i \sum_{\bm{p}} \tilde{\m{G}}^<_{\bm{p}}(\bm{r},t)
&\equiv
-i\sum_{\bm{p}} \int_{-\infty}^\infty \frac{d\omega}{2\pi} \tilde{\m{G}}^{<}(\bm{p}, \omega ,\bm{r},t)
\notag\\[4pt]
&=
\left(\begin{array}{cc}
n_\up(\bm{r},t) & \frac{{\bar \Delta}(\bm{r},t)}{U} \\[6pt]
\frac{{\bar \Delta}^*(\bm{r},t)}{U} & 1-n_\down(\bm{r},t)
\end{array}\right),
\label{eq_N_OP}
\end{align}
Note that the diagonal component is directly related to the particle density $n_\sigma({\bm r},t)$ of the $\sigma$-spin component. Therefore, the order parameter dynamics and the stability of the steady state solutions in the superfluid phase can be examined by analyzing the dynamics of Eq. \eqref{eq_Wigner_trans1}.
\par
Below, we consider the dynamics of $\omega$-integrated (and Wigner-transformed) lesser Green's function, $\bar{\m{G}}^<_{\bm{p}}(\bm{r},t)$. Using the Dyson's equation, we arrive at \cite{Rammer2007, Baym1961} (See Appendix A  for deviation.)
\begin{widetext}
\begin{align}
&i\partial_t \tilde{\m{G}}^<_{\bm{p}}(\bm{r},t) -
\big[\xi_{\bm{p}}\tau_3, \tilde{\m{G}}^<_{\bm{p}}(\bm{r},t) \big]_- -
\left[\frac{{\bm Q}^2}{8m}\tau_3,  \tilde{\m{G}}^<_{\bm{p}}(\bm{r},t) \right]_- 
+\left[\frac{1}{8m}\tau_3, \nabla_{\bm r}^2 \tilde{\m{G}}^<_{\bm{p}}(\bm{r}, t)\right]_-
+ \frac{i}{2}\left[{\bm{p} \over m}\tau_3, \nabla_{\bm r} \tilde{\m{G}}^<_{\bm{p}}(\bm{r}, t) \right]_+ 
\notag\\[6pt]
&\hspace{0.5cm} + \frac{i}{2} \left[ \frac{\bm{Q}}{2m} \tau_0, \nabla _{\bm r}\tilde{\m{G}}^<_{\bm{p}}(\bm{r}, t)\right]_+=\m{I}_{\bm{p}}(\bm{r},t).
\label{EOS_Glesser}
\end{align}
In Eq. (\ref{EOS_Glesser}), the collision term $\m{I}_{\bm{p}}(\bm{r},t)=\m{I}^{\rm int}_{\bm{p}}(\bm{r},t) +\m{I}^{\rm env}_{\bm{p}}(\bm{r},t)$ consists of the interaction term,
\begin{eqnarray}
\m{I}^{\rm int}_{\bm{p}}(\bm{r},t) =\int_{-\infty}^{\infty} \frac{d\omega}{2\pi}\big[ \tilde{\Sigma}^{\rm R}_{\rm int} \circ \tilde{\m{G}}^< -\tilde{\m{G}}^< \circ \tilde{\Sigma}^{\rm A}_{\rm int} +\tilde{\Sigma}^<_{\rm int} \circ \tilde{\m{G}}^{\rm A} -\tilde{\m{G}}^{\rm R} \circ \tilde{\Sigma}^<_{\rm int} \big](\bm{p},\omega,\bm{r},t),
\label{eq.collA}
\end{eqnarray}
as well as the environment term,
\begin{eqnarray}
\m{I}^{\rm env}_{\bm{p}}(\bm{r},t) 
=\int_{-\infty}^{\infty} \frac{d\omega}{2\pi}
\big[ 
\tilde{\Sigma}^{\rm R}_{\rm env} \circ 
\tilde{\m{G}}^< -\m{G}^< \circ 
\tilde{\Sigma}^{\rm A}_{\rm env} 
+\tilde{\Sigma}^<_{\rm env} \circ 
\tilde{\m{G}}^{\rm A} -\tilde{\m{G}}^{\rm R} \circ 
\tilde{\Sigma}^<_{\rm env}  
\big](\bm{p},\omega,\bm{r},t). 
\label{eq.collB}
\end{eqnarray}
In Eqs. (\ref{eq.collA}) and (\ref{eq.collB}), $[A\circ B](\bm{p},\omega,\bm{r},t)$ is the Wigner transformation of the convolution $[A\circ B](1,2)$ in Eq. (\ref{eq.circ}), which is known to be represented by \cite{Rammer2007,Zagoskin2014}
\begin{align}
\bigl[A\circ B\bigr](\bm{p},\omega, \bm{r}, t)
&=
A(\bm{p},\omega, \bm{r}, t)
e^{\frac{i}{2}[\overleftarrow{\partial_\omega} \overrightarrow{\partial_t} -\overleftarrow{\partial_t} \overrightarrow{\partial_\omega}  + \overleftarrow{\partial_{\bm{r}}} \cdot \overrightarrow{\partial_{\bm{p}}} - \overleftarrow{\partial_{\bm{p}}} \cdot \overrightarrow{\partial_{\bm{r}}}]}
B(\bm{p},\omega,\bm{r},t)
\nonumber\\[8pt]
&=
\scalebox{0.98}{$\displaystyle
A(\bm{p},\omega, \bm{r}, t) B(\bm{p},\omega, \bm{r}, t)
+
\frac{i}{2} A(\bm{p},\omega, \bm{r}, t) \left[\overleftarrow{\partial_\omega} \overrightarrow{\partial_t} -\overleftarrow{\partial_t} \overrightarrow{\partial_\omega}  + \overleftarrow{\partial_{\bm{r}}} \cdot \overrightarrow{\partial_{\bm{p}}} - \overleftarrow{\partial_{\bm{p}}} \cdot \overrightarrow{\partial_{\bm{r}}} \right] B(\bm{p},\omega, \bm{r}, t) + \cdots. $}
\label{eq_Moyal}
\end{align}
\end{widetext}
Here, the left (right) arrow on each differential operator means that it acts on the left (right) side of this operator. Since we are interested in the slowly varying dynamics both in terms of time and space, below, we will only retain up to the first order with respect to $\left[\overleftarrow{\partial_\omega} \overrightarrow{\partial_t} -\overleftarrow{\partial_t} \overrightarrow{\partial_\omega}  + \overleftarrow{\partial_{\bm{r}}} \cdot \overrightarrow{\partial_{\bm{p}}} - \overleftarrow{\partial_{\bm{p}}} \cdot \overrightarrow{\partial_{\bm{r}}} \right]$, which greatly simplifies the analysis.
\par
Let us first evaluate further the interaction term $\m{I}^{\rm int}_{\bm{p}}(\bm{r},t)$ of the collision term in Eq. (\ref{eq.collA}). The self-energies ${\tilde \Sigma}_{\rm int}^{\rm R,A,<}$ in the Wigner representation that enters $\m{I}^{\rm int}_{\bm{p}}(\bm{r},t)$ also depend on $t$ through $\bar{\Delta}({\bm r},t)$ as
\begin{align}
\tilde{\Sigma}^{\rm R}_{\rm int}(\bm{p}, \omega, \bm{r}, t) &= -\big[{\bar \Delta}(\bm{r},t)\tau_+ +{\bar \Delta}^*(\bm{r},t) \tau_-\big] s
\equiv -\tilde{\Delta}(\bm{r},t) 
\label{eq_Wigner_self_int_R},
\\
\tilde{\Sigma}^{\rm A}_{\rm int}(\bm{p}, \omega, \bm{r}, t) &= -{\tilde \Delta}^\dagger({\bm r},t),
\label{eq_Wigner_self_int_A}
\\[4pt]
\tilde{\Sigma}^<_{\rm int}(\bm{p},\omega,\bm{r},t) &= 0. 
\label{eq_Wigner_self_int_L}
\end{align}
Within the first order gradient approximation of the Moyal product mentioned above, Eq. (\ref{eq.collA}) is evaluated as
\begin{widetext}
\begin{align}
\m{I}_{\bm{p}}^{\rm int}(\bm{r}, t) 
&\simeq
-\big[ \tilde{\Delta}(\bm{r},t), \tilde{\m{G}}^<_{\bm{p}}(\bm{r},t) \big]_-
-\frac{i}{2} \big[\nabla_{\bm r} \tilde{\Delta}(\bm{r},t), \partial_{\bm{p}} \tilde{\m{G}}^<_{\bm{p}}(\bm{r},t) \big]_+ 
+ \frac{i}{2} \int_{-\infty}^{\infty} \frac{d\omega}{2\pi} 
\big[\partial_t \tilde{\Delta}(\bm{r},t), 
\partial_\omega \tilde{\m{G}}^<(\bm{p},\omega, \bm{r},t) 
\big]_+ 
\notag\\[6pt]
&=
-\big[ \tilde{\Delta}(\bm{r},t), \tilde{\m{G}}^<_{\bm{p}}(\bm{r},t) \big]_- 
-\frac{i}{2} \big[ \nabla_{\bm r} \tilde{\Delta}(\bm{r},t),  \nabla_{\bm{p}} \tilde{\m{G}}^<_{\bm{p}}(\bm{r},t) \big]_+, 
\label{eq_Iint}
\end{align}
\end{widetext}
where the last term in the first expression vanishes by the integration by parts. 
\par
For the environment part $\m{I}_{\bm{p}}^{\rm env}(\bm{r}, t)$ of the collision term in Eq. (\ref{eq.collB}), we take advantage of the fact that the reservoirs are assumed to be huge compared to the main system and remain unchanged.
Because of this property, the self energy ${\hat {\tilde \Sigma}}_{\rm env}({\bm p},\omega)$ that enters $\m{I}_{\bm{p}}^{\rm env}(\bm{r}, t)$ can be expressed as,
\begin{align}
\tilde{\Sigma}^{\rm R}_{\rm env}(\bm{p}, \omega, \bm{r}, t) &= -2i\gamma \tau_0 
\label{eq_Wigner_self_env_R},
\\[4pt]
\tilde{\Sigma}^{\rm A}_{\rm env}(\bm{p}, \omega, \bm{r}, t) &= 2i\gamma \tau_0 \label{eq_Wigner_self_env_A},
\\[4pt]
\tilde{\Sigma}^<_{\rm env}(\bm{p}, \omega, \bm{r}, t) &= 4i\gamma F(\omega) \tau_0 
\label{eq_Wigner_self_env_L},
\end{align}
which are actually the same as the steady-state counterpart given in Eq. (\ref{self_tilde_env}). Substituting these into Eq. (\ref{eq.collB}), we obtain, again within the first-order gradient expansion,
\begin{align}
\m{I}_{\bm{p}}^{\rm env}(\bm{r}, t) 
& \simeq 
-4i\gamma \tilde{\m{G}}^<_{\bm{p}}(\bm{r},t) -4\gamma\int_{-\infty}^\infty \frac{d\omega}{2\pi} F(\omega) \m{A}(\bm{p},\omega,\bm{r},t) 
\notag\\[4pt]
&\hspace{0.7cm}
+4\gamma \int_{-\infty}^{\infty} \frac{d\omega}{2\pi} \partial_\omega F(\omega) \partial_t \m{R}(\bm{p},\omega ,\bm{r},t),
\label{eq_Ienv_Wig}
\end{align}
where we have introduced the spectral weight \cite{Rammer2007},
\begin{equation}
\m{A}(\bm{p}, \omega, \bm{r}, t) = i \big[\tilde{\m{G}}^{\rm R} -\tilde{\m{G}}^{\rm A}\big](\bm{p}, \omega, \bm{r}, t), 
\label{eq_SW}
\end{equation}
and 
\begin{equation}
\m{R}(\bm{p}, \omega, \bm{r}, t) = \frac{1}{2} \big[\tilde{\m{G}}^{\rm R} +\tilde{\m{G}}^{\rm A}\big](\bm{p}, \omega, \bm{r}, t).
\label{eq.R}
\end{equation}
As shown in Appendix B, the last term in Eq. (\ref{eq_Ienv_Wig}) vanishes in the present case. Thus, we obtain
\begin{equation}
\m{I}_{\bm{p}}^{\rm env}(\bm{r}, t) 
= -4i\gamma \tilde{\m{G}}^<_{\bm{p}}(\bm{r},t) -4\gamma\int_{-\infty}^\infty \frac{d\omega}{2\pi} F(\omega) \m{A}(\bm{p},\omega,\bm{r},t).
\label{eq_Ienv_Wig2}
\end{equation}
Using the expressions for $\tilde{\m{G}}^{\rm R}(\bm{p},\omega,\bm{r},t)$ and $\tilde{\m{G}}^{\rm A}(\bm{p},\omega,\bm{r},t)$ given, respectively, in Eqs. (\ref{eq_app_GR}) and (\ref{eq_app_GA}), one finds that the time-dependent spectral function $\m{A}(\bm{p},\omega,\bm{r},t)$ in Eq. (\ref{eq_SW}) has the form,
\begin{equation}
\m{A}(\bm{p},\omega,\bm{r},t) =
\sum_{\eta=\pm} \frac{4\gamma}{\big[\omega -E^\eta_{\bm{p},\bm{Q}}(\bm{r},t)\big]^2 +4\gamma^2} \Xi^\eta_{\bm{p},\bm{Q}}(\bm{r},t), 
\label{eq_TDSW}
\end{equation}
where $E^\pm_{\bm{p},\bm{Q}}(\bm{r},t)$ and $\Xi^\eta_{\bm{p},\bm{Q}}(\bm{r},t)$ are given in Eqs. (\ref{app_E}) and (\ref{app_M}), respectively.
\par
Substituting Eqs. (\ref{eq_Iint}) and (\ref{eq_Ienv_Wig2}) into Eq. (\ref{EOS_Glesser}), we obtain the desired QKE,
\begin{widetext}
\begin{align}
&
i\partial_t \tilde{\m{G}}^<_{\bm{p}}(\bm{r},t)  
= 
\big[
\xi_{\bm{p}} \tau_3 - \tilde{\Delta}(\bm{r},t),  
\tilde{\m{G}}^<_{\bm{p}}(\bm{r},t)
\big]_- 
+
\left[ 
\frac{{\bm Q}^2}{8m}\tau_3,\tilde{\m{G}}^<_{\bm{p}}(\bm{r},t) 
\right]_- 
-
\left[ 
\frac{1}{8m} \tau_3,\nabla_{\bm r}^2 \tilde{\m{G}}^<_{\bm{p}}(\bm{r},t) 
\right]_- 
-
\frac{i}{2} 
\left[
{{\bm p} \over m} \tau_3 , \nabla_{\bm r} \tilde{\m{G}}^<_{\bm{p}}(\bm{r},t) 
\right]_+
\nonumber\\[6pt]
&
\hspace{1cm}
-
\frac{i}{2} 
\left[
\frac{\bm{Q}}{2m} \tau_0, \nabla_{\bm r} \tilde{\m{G}}^<_{\bm{p}}(\bm{r},t)
\right]_+
-
4i\gamma \tilde{\m{G}}^<_{\bm{p}}(\bm{r},t)
-
4\gamma \int_{-\infty}^\infty 
\frac{d\omega}{2\pi} F(\omega) \m{A}(\bm{p},\omega,\bm{r},t) 
-
\frac{i}{2} 
\big[
\nabla_{\bm r} \tilde{\Delta}(\bm{r},t), 
\nabla_{\bm{p}} \tilde{\m{G}}^<_{\bm{p}}(\bm{r},t) 
\big]_+.
\label{eq_kinetic}
\end{align}
\end{widetext}
In fact, the last term does not actually affect the time evolution of the superfluid order parameter ${\bar \Delta}(\bm{r},t)$. To see this, we recall that ${\bar \Delta}(\bm{r},t)$ is related to the (gauge-transformed) lesser Green's function as
\begin{equation}
{\bar \Delta}(\bm{r},t) = -iU \sum_{\bm{p}} \tilde{\m{G}}^<_{\bm{p}}(\bm{r},t)_{12}.
\label{eq_OP_fluc_A}
\end{equation}
Thus, the equation of motion for the superfluid order parameter ${\bar \Delta}(\bm{r},t)$ is obtained from the $\bm{p}$-summation of the (12)-component of Eq. (\ref{eq_kinetic}). In the resulting equation, the contribution coming from the last term in Eq. (\ref{eq_kinetic}) vanishes. Since our stability analysis only needs the time evolution of $\bar{\Delta}({\bm r},t)$, the last term in Eq. (\ref{eq_kinetic}) may be ignored for our purpose \cite{note2}.
\par
We also note that the first term on the right hand side in Eq. \eqref{eq_kinetic} represents the unitary time evolution. When we only retain this term and further assume a uniform superfluid $\bar{\Delta}(\bm{r},t) = \bar{\Delta}(t)$, the QKE (\ref{eq_kinetic}) is reduced to
\begin{equation}
i\partial_t \tilde{\m{G}}^<_{\bm{p}}(t) = \big[\xi_{\bm{p}} \tau_3 -\tilde{\Delta}(t) , \tilde{\m{G}}^<_{\bm{p}}(t) \big]. 
\label{eq_TDBdG}
\end{equation}
This is equivalent to the so-called time-dependent Bogoliubov-de Gennes (TDBdG) equation \cite{Ketterson1998,Andreev1964}, which has widely been used in studying the dynamics of a closed Fermi condensate. In this sense, Eq. (\ref{eq_kinetic}) may be interpreted as an extension of TDBdG theory to the open Fermi system shown in Fig. \ref{fig1}(a).
\par
%%%%%%%%%%%%%%%%%%%%%%%%%%%%%%%%%%%%%%%%%%%%%%%%%%%%%%%%%%%%%%%%%%%%%%%%%%%%%%%%
\par
\subsection{Stability analysis of non-equilibrium superfluid steady states}
\par
Using the QKE scheme discussed in Sec. III.A, we are now in the position to study the stability of the obtained steady states. We compute the time evolution of the superfluid order parameter $\bar\Delta(\bm r,t)$ in the situation where the initial condition is prepared arbitrarily close to the steady-state solution $\Delta_0$. We can then judge the stability of the solution by checking whether the deviation converges to zero or amplifies even more. It is useful to consider the deviation of the superfluid order parameter from the steady-state value:
\begin{equation}
\delta \bar{\Delta}(\bm{r},t)= \bar{\Delta}(\bm{r},t) -\Delta_0 = -i U\sum_{\bm{p}} \delta \tilde{\m{G}}^<_{\bm{p}}(\bm{r},t)_{12},
\label{eq_delta_Delta}
\end{equation}
Here, $\delta \tilde{\m{G}}^<_{\bm{p}}(\bm{r},t) \equiv \tilde{\m{G}}^<_{\bm{p}}(\bm{r},t) -\tilde{\m{G}}^<_{\bm{p}, {\rm NESS}}$ is the deviation of the lesser Green's function from the (non-equilibrium) mean-field value, where
\begin{equation}
{\tilde {\m G}}^<_{{\bm p},{\rm NESS}}=
\int{d\omega \over 2\pi} {\tilde {\m G}}^<_{{\bm p},{\rm NESS}}(\omega)
\end{equation}
is the $\omega$-integrated lesser Green's function in the non-equilibrium steady state, with $\tilde{\m{G}}^<(\bm{p},\omega)$ being given in Eq. (\ref{eq_NESS_GL}). $\delta\bar\Delta(\bm r,t)$ decays (amplifies) as a function of time if the steady state solution is (un)stable. 
\par
To derive the equation for the dynamics of $\delta\bar\Delta(\bm r,t)$, it is useful to linearize the QKE (\ref{eq_kinetic}) in terms of $\delta \bar{\Delta}(\bm{r},t)$ and $\delta \tilde{\m{G}}^<_{\bm{p}}(\bm{r},t)$. Carrying out the Fourier transformation with respect to $\bm{r}$, one has
\begin{widetext}
\begin{align}
i\partial_t \delta \tilde{\m{G}}^<_{\bm{p}}(\bm{q},t)=& 
\big[\xi_{\bm{p}} \tau_3 - \tilde{\Delta}_0,  \delta \tilde{\m{G}}^<_{\bm{p}}(\bm{q},t)\big]_- 
-\big[\delta\tilde{\Delta}(\bm{q},t),  \tilde{\m{G}}^<_{\bm{p},{\rm NESS}} \big]_- 
-\frac{Q^2 -q^2}{8m}\left[ \tau_3, \delta\tilde{\m{G}}^<_{\bm{p}}(\bm{q},t) \right]_- 
\nonumber\\[4pt]
&+
\frac{\bm{p}\cdot \bm{q}}{2m}
\left[
\tau_3 , \delta\tilde{\m{G}}^<_{\bm{p}}(\bm{q},t) 
\right]_+
+\frac{\bm{Q}\cdot \bm{q}}{4m}
\left[
\tau_0 , \delta\tilde{\m{G}}^<_{\bm{p}}(\bm{q},t) 
\right]_+
-
4i\gamma \delta\tilde{\m{G}}^<_{\bm{p}}(\bm{r},t)
-
4\gamma \int_{-\infty}^\infty 
\frac{d\omega}{2\pi} F(\omega) \delta\m{A}(\bm{p},\omega,\bm{q},t),
\label{eq_linearized_QKE}
\end{align}
where ${\tilde \Delta}_0=\Delta_0\tau_1$ and
\begin{equation}
\delta\tilde{\m{G}}^<_{\bm{p}}(\bm{q},t)=\int d{\bm r}e^{-i{\bm q}\cdot{\bm r}}
\delta\tilde{\m{G}}^<_{\bm{p}}(\bm{r},t),
\label{eq.FGL}
\end{equation}
\begin{equation}
\delta\tilde{\Delta}(\bm{q},t)=\int d{\bm r}e^{-i{\bm q}\cdot{\bm r}}
\delta\tilde{\Delta}(\bm{r},t)
\equiv 
\delta\bar{\Delta}(\bm{q},t) \tau_+ +\delta\bar{\Delta}^*(\bm{q},t)\tau_-.
\label{eq.FDELTA}
\end{equation}
Here,
\begin{equation}
\delta\m{A}(\bm{p},\omega, \bm{q},t)
=
\sum_{\eta=\pm} \bigg[
\frac{4\gamma}{[\omega -\eta E^{0,\eta}_{\bm{p},\bm{Q}}]^2 +4\gamma^2} \delta \Xi^\eta_{\bm{p},\bm{Q}}(\bm{q},t) 
+
\frac{8\gamma\big[\omega -\eta E^{0,\eta}_{\bm{p},\bm{Q}} \big]}{\big[[\omega -\eta E^{0,\eta}_{\bm{p},\bm{Q}}]^2 +4\gamma^2\big]^2} \Xi^{0,\eta}_{\bm{p},\bm{Q}} \delta E_{\bm{p},\bm{Q}}(\bm{q},t) \bigg]
\end{equation}
is the linearized time-dependent spectral function, which is obtained by linearizing $\m{A}(\bm{p},\omega,\bm{r},t)$ in Eq. (\ref{eq_TDSW}) with respect to $\delta \bar{\Delta}(\bm{r},t)$. We have also introduced here 
\begin{equation}
\delta \Xi^{+}_{\bm{p},\bm{Q}}(\bm{q},t) = -\delta \Xi^{-}_{\bm{p},\bm{Q}}(\bm{q},t) \notag\\ 
= -\frac{1}{2\big[E_{\bm{p},\bm{Q}}^0 \big]^2}
\left(
\begin{array}{cc}
\xi^{\rm (s)}_{\bm{p},\bm{Q}} \delta E_{\bm{p},\bm{Q}}(\bm{q},t) & 
E^0_{\bm{p},\bm{Q}}\delta \bar{\Delta}(\bm{q},t) +\Delta_0 \delta E_{\bm{p},\bm{Q}}(\bm{q},t) \\[4pt]
E^0_{\bm{p},\bm{Q}}\delta \bar{\Delta}^*(\bm{q},t) +\Delta_0 \delta E_{\bm{p},\bm{Q}}(\bm{q},t) &
-\xi^{\rm (s)}_{\bm{p},\bm{Q}} \delta E_{\bm{p},\bm{Q}}(\bm{q},t)
\end{array}
\right),
\end{equation}
\end{widetext}
and
\begin{equation}
\delta E_{\bm{p},\bm{Q}}(\bm{q},t) = \frac{\Delta_0}{E^0_{\bm{p},\bm{Q}}} {\rm Re}\big[\delta \bar{\Delta}(\bm{q},t) \big].
\end{equation}
\par
Since Eq. (\ref{eq_linearized_QKE}) does not involve any mode-coupling term, one may safely focus on a particular value of the momentum ${\bm q}$ ($\equiv{\bar {\bm q}}$) in considering how the initial deviation of the superfluid order parameter from the mean-field value evolves over time. Keeping this in mind, we take the following initial condition for the quantum kinetic equation (\ref{eq_linearized_QKE}): 
\begin{equation}
\delta \bar{\Delta}({\bar {\bm q}}, t=0) = \big[\Delta_0 +\delta|\bar{\Delta}
({\bar {\bm q}},t=0)| \big] e^{i{\bar {\bm q}}\cdot \bm{r}} -\Delta_0.
\label{eq_initial}
\end{equation}
Here, $\delta|\bar{\Delta}({\bar {\bm q}},t=0)|$ and ${\bar {\bm q}}\cdot{\bm r}$ physically have the meanings of amplitude and phase deviations from the mean-field value $\Delta_0$, respectively. 
\par
We numerically solve Eq. (\ref{eq_linearized_QKE}) in the weak-coupling regime $(p_{\rm F}a_s)^{-1}=-1$, by using the fourth order implicit Runge-Kutta method with small time steps. At each time step, the deviation $\delta\bar{\Delta}({\bar {\bm q}}, t)$ in the right-hand-side of this equation is evaluated from $\delta\tilde{\m{G}}^<_{\bm{p}}({\bar {\bm q}},t)$, to proceed to the next time step. Because our QKE approach uses the gradient expansion as explained in Sec. III A, we set the initial condition so as to satisfy $\delta|\bar{\Delta}({\bar {\bm q}},t=0)|\ll \Delta_0$ and $|{\bar {\bm q}}| \ll p_{\rm F}$ \cite{note3, KopninBook}. 
\par
%%%%%%%%%%%%%%%%%%%%%%%%%%%%%%%%%%%%%%%%%%%%%%%%%%%%%%%%%%%%%%%%%%%%%%%%%%%%%%%%
\begin{figure}[tb]
\centering
\includegraphics[width=7.8cm]{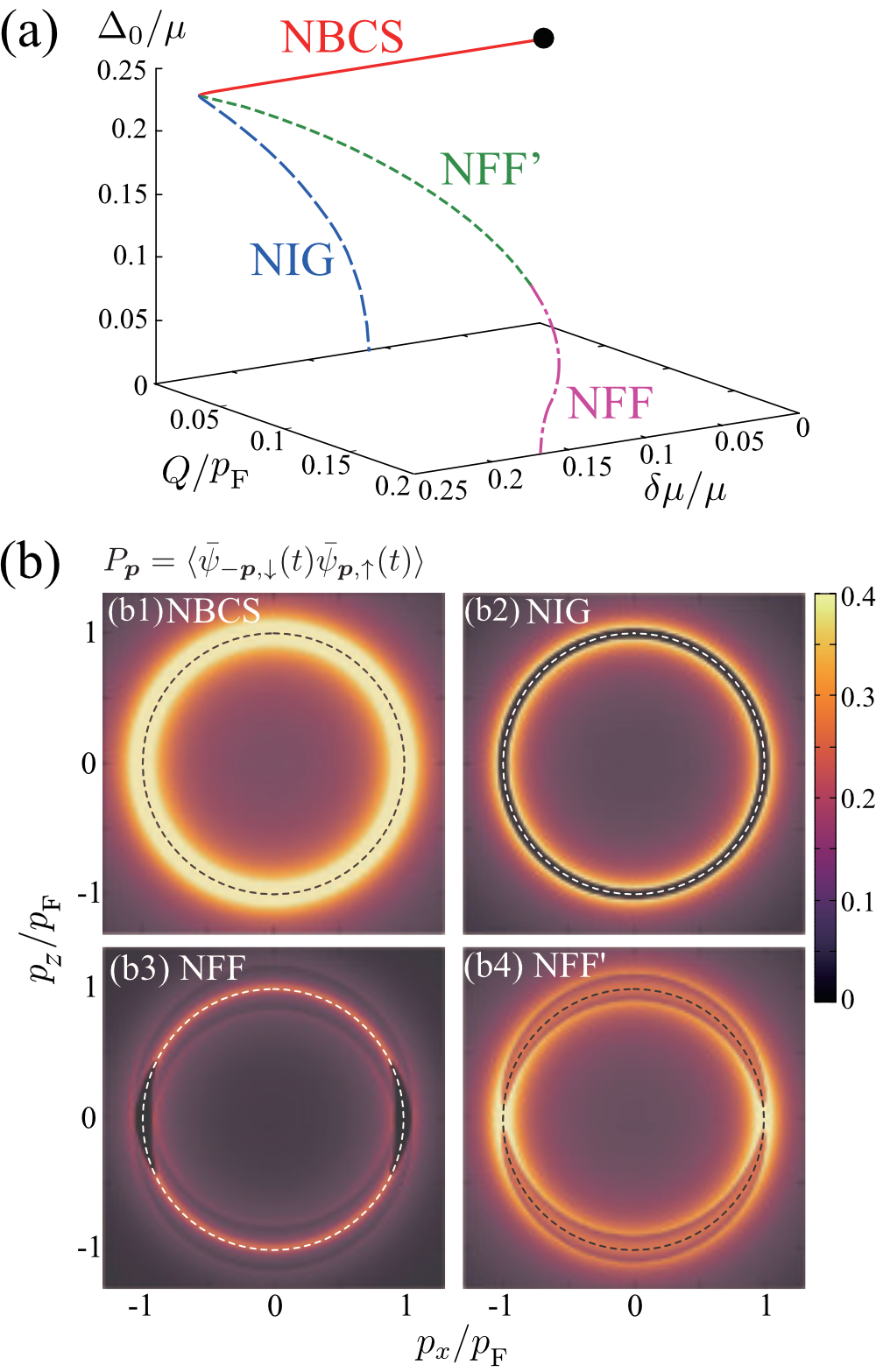}
\caption{(a) Non-equilibrium superfluid solutions of the gap equation (\ref{eq_NESS_gap}) in the weak-coupling regime ($(p_{\rm F}a_a)^{-1}=-1$) of a driven-dissipative Fermi gas. We set $T_{\rm env}=0$ and $\gamma\to +0$, and impose the vanishing current condition in Eq. (\ref{eq_NESS_current}). Among the four mean-field solutions, NBCS and NIG (non-equilibrium interior gap state) are uniform superfluid states ($\bm{Q}=0$). NFF and NFF' are FF-like non-uniform states (${\bm Q}\ne 0$ and ${\bm Q}\parallel p_z$). The solid circle is at the BCS state in the thermal equilibrium case ($\delta\mu=0$). (b) Calculated intensity of the pair amplitude of each state, when $\delta\mu=0.145\mu$. The dotted line in each panel shows the position at $p=p_{\rm F}=\sqrt{2m\mu}$. We will show in Sec. \ref{subsec_stability} that NBCS and NFF are stable solutions, while NIG and NFF' are unstable solutions.}
\label{fig_NESS} 
\end{figure}
%%%%%%%%%%%%%%%%%%%%%%%%%%%%%%%%%%%%%%%%%%%%%%%%%%%%%%%%%%%%%%%%%%%%%%%%%%%%%%%
\par
\section{Non-equilibrium superfluid steady states in driven-dissipative Fermi gas}
\par
We now explore non-equilibrium superfluid steady states in a driven-dissipative Fermi gas. In Sec. IV.A, we first look for possible mean-field solutions of the gap equation (\ref{eq_NESS_gap}), under the vanishing current condition in Eq. (\ref{eq_NESS_current}).  As we have emphasized in the previous section, the solution of the gap equation (\ref{eq_NESS_gap}) does not necessarily mean the stability of this pairing state. To assess which solutions are physical, we apply the stability analysis explained in Sec. III.B to each mean-field solution in Sec. IV.B.
\par
%%%%%%%%%%%%%%%%%%%%%%%%%%%%%%%%%%%%%%%%%%%%%%%%%%%%%%%%%%%%%%%%%%%%%%%%%%%%%%%%
\par
\subsection{Non-equilibrium superfluid steady states}
\par
Figure \ref{fig_NESS}(a) summarizes the non-equilibrium superfluid steady state solutions of the gap equation (\ref{eq_NESS_gap}) under the vanishing current condition in Eq. (\ref{eq_NESS_current}), in the weak-coupling regime ($(p_{\rm F}a_s)^{-1}=-1$) of a driven-dissipative Fermi gas at $T_{\rm env}=0$ and $\gamma\to +0$ \cite{noteGamma}. As seen in this figure, four self-consistent solutions are obtained under the ansatz in Eq. (\ref{eq_OP}). Among them, NBCS (non-equilibrium BCS state) and NIG (non-equilibrium interior gap state) are isotropic superfluid states (${\bm Q}=0$). In particular, NBCS is reduced to the ordinary BCS state in the thermal equilibrium limit $\delta\mu\to0$ (solid circle in Fig. \ref{fig_NESS}(a)), so that it may be viewed as an extension of the ordinary BCS state to the non-equilibrium steady state. On the other hand, NIG is similar to the so-called interior gap state \cite{Sarma1963, Liu2003} that arises in spin-imbalanced system, as we argue in the following.  In the next subsection, we will show further that NBCS (NIG) is a (un)stable state, which is in parallel to the known results in equilibrium that the ordinary BCS state (interior gap state) is (un)stable against superfluid fluctuations (See Table I.). The remaining two solutions, NFF and NFF' (non-equilibrium Fulde-Ferrell states), are anisotropic superfluid state with ${\bm Q}\ne 0$. We briefly note that the existence of two kinds of FF states has also been pointed out in a thermal-equilibrium spin-imbalanced Fermi gas \cite{Hu2006}. Actually, the former (latter) turns out to be a (un)stable state, as we will discuss in the next subsection, which is again parallel to the equilibrium case (See Table I.).
\par
To grasp the character of each state, we conveniently consider the pair amplitude,
\begin{equation}
P_{\bm p}\equiv \langle{\bar \psi}_{-{\bm p},\downarrow}(t){\bar \psi}_{{\bm p},\uparrow}(t)\rangle,
\label{eq.PA_def}
\end{equation} 
which physically describes the pairing structure in momentum space. Here, 
\begin{equation}
{\bar \psi}_{{\bm p},\sigma}(t)=\int d{\bm r} e^{-i{\bm p}\cdot{\bm r}}{\bar \psi}_\sigma({\bm r},t)
\end{equation}
is the Fourier transformation of the gauge-transformed field operator,
\begin{equation}
{\bar \psi}_\sigma({\bm r},t)=e^{-i\chi({\bm r},t)}\psi_\sigma({\bm r},t),
\end{equation}
where the phase $\chi$ is given in Eq. (\ref{eq.gaugeC}). Within the present mean-field scheme, $P_{\bm p}$ in Eq. (\ref{eq.PA_def}) can be evaluated from the lesser Green's function ${\tilde {\m G}}^<({\bm p},\omega)$ in Eq. (\ref{eq_NESS_GL}) as
\begin{align}
P_{\bm p}
&=
-i\int{d\omega \over 2\pi}{\tilde {\m G}}^<_{12}({\bm p},\omega)
\notag\\[4pt]
&=
-{\Delta_0 \over E_{{\bm p},{\bm Q}}^0}
\sum_{\eta=\pm}\eta \int {d\omega \over 2\pi}
F(\omega){2\gamma \over [\omega-\eta E_{{\bm p},{\bm Q}}^{0,\eta}]^2+4\gamma^2}.
\label{eq.PA}
\end{align}
In the small damping limit $\gamma\to +0$ (which is the case of Fig. \ref{fig_NESS}(a)), one may carry out the $\omega$-integration in Eq. (\ref{eq.PA}), giving
\begin{eqnarray}
P_{\bm p}=
{\Delta_0 \over 2E_{{\bm p},{\bm Q}}^0}
\left[
F(-E_{{\bm p},{\bm Q}}^{0,-})-F(E_{{\bm p},{\bm Q}}^{0,+})
\right].
\label{eq.PA200}
\end{eqnarray}
\par
In the NBCS and NIG cases (${\bm Q}=0$), Eq. (\ref{eq.PA200}) is reduced to
\begin{equation}
P_{\bm p}=
{\Delta_0 \over 2E_{\bm p}}[1-2F(E_{\bm p})],
\end{equation}
where $E_{\bm p}$ is given below Eq. (\ref{eq.BCSgap}). In the thermal equilibrium BCS limit ($\delta\mu\to 0$), $F(E_{\bm p})=f(E_{\bm p})$ vanishes at $T_{\rm env}=0$. Then, as well-known in the ordinary BCS theory, the pair amplitude,
\begin{equation}
P_{\bm p}=
{\Delta_0 \over 2\sqrt{(\varepsilon_{\bm p}-\mu)^2+\Delta_0^2}},
\label{eq.PA_BCS}
\end{equation}
has large intensity around the Fermi momentum $p_{\rm F}=\sqrt{2m\mu}$, indicating that Cooper pairs are dominantly formed around the Fermi surface. We also find from the definition of $F(\omega)$ in Eq. (\ref{eq.F}) that, even when $\delta\mu>0$, Eq. (\ref{eq.PA_BCS}) still holds, as far as $\Delta_0\ge\delta\mu$. This is just the NBCS case. Indeed, as shown in Fig. \ref{fig_NESS}(b1), the calculated NBCS pair amplitude has large intensity around $p_{\rm F}=\sqrt{2m\mu}$ (although $p_{\rm F}$ no longer has the meaning of the Fermi momentum, when $\delta\mu=0.145\mu>0$). 
\par
%%%%%%%%%%%%%%%%%%%%%%%%%%%%%%%%%%%%%%%%%%%%%%%%%%%%%%%%%%%%%%%%%%%%%%%%%%%%%%%%
\begin{table}[bt]
\caption{Summary of the stability analysis for $T_{\rm env}=0$ and $\gamma=0.005\mu$. Among the superfluid solutions, the underlined states are only stable.}
\centering
\begin{tabular}{ll| l } \hline\hline
\multicolumn{2}{c|}{region} & \multicolumn{1}{c}{superfluid solutions}  \\\hline
region I &($0<\delta\mu\le 0.111\mu$) & \underline{NBCS}  \\
region II &($0.111\mu<\delta\mu\le 0.135\mu$) & \underline{NBCS}, NIG \\
region III &($0.135\mu<\delta\mu\le 0.152\mu$) &\underline{NBCS}, NIG, \underline{NFF}, NFF'  \\
region IV &($0.152\mu<\delta\mu\le 0.183\mu$)& \underline{NBCS}, NIG, NFF' \\ \hline\hline
\end{tabular} 
\label{table1}
\end{table}%
%%%%%%%%%%%%%%%%%%%%%%%%%%%%%%%%%%%%%%%%%%%%%%%%%%%%%%%%%%%%%%%%%%%%%%%%%%%%%%%
\par
For NIG, although the pair amplitude $P_{\bm p}$ is also isotropic, it vanishes around the ``Fermi momentum" $p=p_{\rm F}$ (see Fig. \ref{fig_NESS}(b2)). In this pairing state, the superfluid order parameter $\Delta_0$ is not so large as the NBCS case (see Fig. \ref{fig_NESS}(a)). When $\Delta_0<\delta\mu$, one obtains
\begin{equation}
P_{\bm p}=
{\Delta_0 \over 2\sqrt{(\varepsilon_{\bm p}-\mu)^2+\Delta_0^2}}
\Theta\left(\sqrt{(\varepsilon_{\bm p}-\mu)^2+\Delta_0^2}-\delta\mu\right).
\label{eq.PA_IG}
\end{equation}
Equation (\ref{eq.PA_IG}) immediately explains the vanishing pairing amplitude around $p=p_{\rm F}$ seen in Fig. \ref{fig_NESS}(b2) (because the step function vanishes there). In addition, the region where $P_{\bm p}>0$ is given by $p\le {\tilde p}_{\rm F1},~{\tilde p}_{\rm F2}\le p$, where
\begin{eqnarray}
{\tilde p}_{\rm F1}&=&\sqrt{2m\left[\mu-\sqrt{\delta\mu^2-\Delta_0^2}\right]},\\
{\tilde p}_{\rm F2}&=&\sqrt{2m\left[\mu+\sqrt{\delta\mu^2-\Delta_0^2}\right]}.
\label{eq.PA_IGregion}
\end{eqnarray}
Particularly in the limiting case, $\delta\mu\gg\Delta_0$, one finds
\begin{eqnarray}
{\tilde p}_{\rm F1}\simeq \sqrt{2m[\mu-\delta\mu]}=p_{\rm F1},\\
{\tilde p}_{\rm F2}\simeq \sqrt{2m[\mu+\delta\mu]}=p_{\rm F2},
\label{eq.PA_IGregion2}
\end{eqnarray}
that coincide with the positions of two edges imprinted on the Fermi momentum distribution $n_{{\bm p},\sigma}$ by the two reservoirs \cite{Kawamura2020} (see Fig. \ref{fig1}). When we simply regard these edges as two `Fermi surfaces' with different sizes, Fig. \ref{fig_NESS}(b2) indicates that NIG Cooper pairs are formed around the `Fermi surfaces' at $p_{\rm F1}$ and $p_{\rm F2}$. This pairing structure is similar to the Sarma(-Liu-Wilczek) state \cite{Sarma1963, Liu2003} (which is also referred to as the interior gap state in the literature) discussed in thermal equilibrium superconductivity under an external magnetic field, as well as in a spin imbalanced Fermi gas.
\par
%%%%%%%%%%%%%%%%%%%%%%%%%%%%%%%%%%%%%%%%%%%%%%%%%%%%%%%%%%%%%%%%%%%%%%%%%%%%%%%%
\begin{figure}[tb]
\centering
\includegraphics[width=7.8cm]{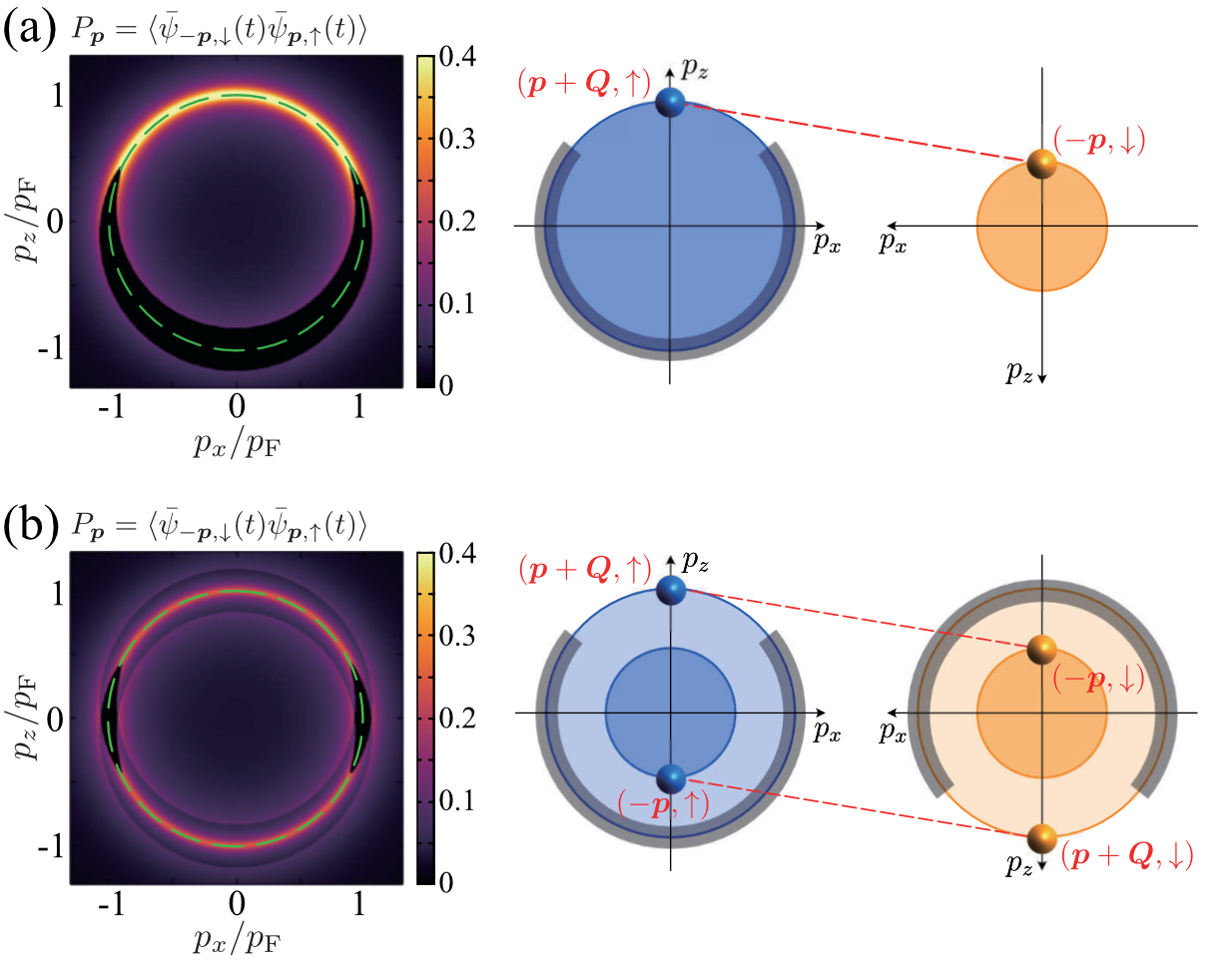}
\caption{
Calculated intensity of the pair amplitude and schematic pictures of pair-formation: (a) FF state. (b) NFF state. Fermions in the shaded regions (blocking regions) do not contribute to the pair formation.}
\label{fig_NFF_FF} 
\end{figure}
%%%%%%%%%%%%%%%%%%%%%%%%%%%%%%%%%%%%%%%%%%%%%%%%%%%%%%%%%%%%%%%%%%%%%%%%%%%%%%%
\par
We next consider the anisotropic NFF and NFF' states with ${\bm Q}\ne 0$. To grasp their pairing structures, it is useful to rewrite Eq. (\ref{eq.PA200}) in the form,
\begin{eqnarray}
P_{\bm p}={1 \over 2}
\left[
P_{\bm p}^{\rm FF}({\bm Q},\delta\mu)+P_{-{\bm p}}^{\rm FF}({\bm Q},-\delta\mu)
\right],
\label{eq.PA201}
\end{eqnarray}
where
\begin{equation}
P_{\bm p}^{\rm FF}({\bm Q},\delta\mu)=
{\Delta_0 \over 2E_{{\bm p},{\bm Q}}^0}
\left[1-f\big({\bar E}^+_{{\bm p},{\bm Q}}(\delta\mu)\big) -f\big({\bar E}^-_{{\bm p},{\bm Q}}(\delta\mu)\big)\right].
\label{eq.PFF}
\end{equation}
In Eq. (\ref{eq.PFF}), ${\bar E}^\pm_{{\bm p},{\bm Q}}(\delta\mu)$ is given in Eq. (\ref{eq_qp_exc}) where $\xi_{\pm{\bm p}+{\bm Q}/2}$ in $\xi_{{\bm p},{\bm Q}}^{(a)}$ is replaced by 
\begin{equation}
{\bar \xi}_{\pm{\bm p}+{\bm Q}/2}
=\varepsilon_{\pm{\bm p}+{\bm Q}/2}-\mu-\delta\mu.
\end{equation}
Equation (\ref{eq.PFF}) is just the same form as the pair amplitude in the thermal equilibrium FF state \cite{Takada1969,Shimahara1994}, when one regards $\delta\mu$ as an {\it external magnetic field}. Thus, the first (second) term in Eq. (\ref{eq.PA201}) may be viewed as the pair amplitude in the FF state under an external magnetic field $\delta\mu$ ($-\delta\mu$), which just corresponds to the pairing (A) ((B)) in Fig. \ref{fig1}(b).
\par
%%%%%%%%%%%%%%%%%%%%%%%%%%%%%%%%%%%%%%%%%%%%%%%%%%%%%%%%%%%%%%%%%%%%%%%%%%%%%%%%
\begin{figure}[tb]
\centering
\includegraphics[width=7.8cm]{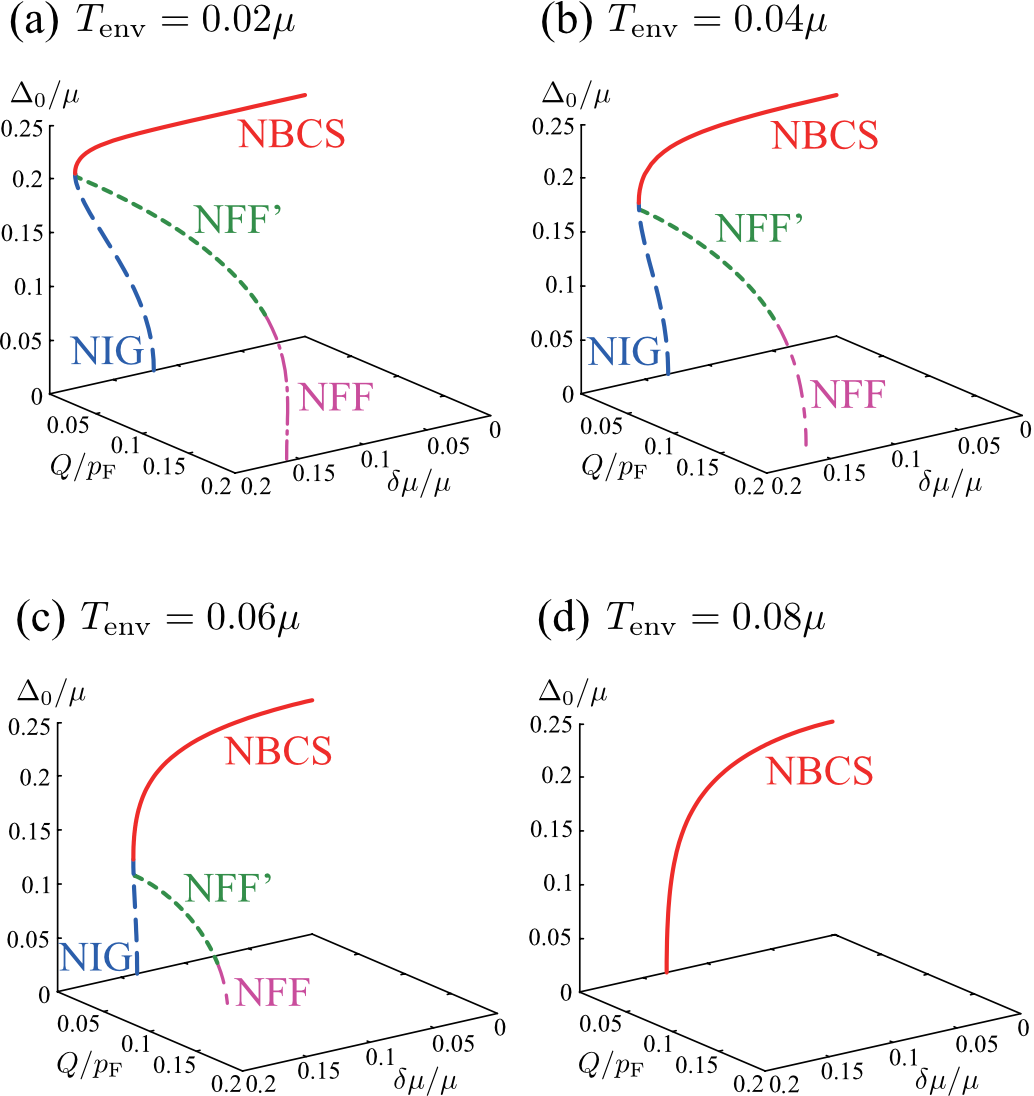}
\caption{Same plots as Fig. \ref{fig_NESS}(a) for non-zero environment temperatures $T_{\rm env}$.}
\label{fig_finiteT} 
\end{figure}
%%%%%%%%%%%%%%%%%%%%%%%%%%%%%%%%%%%%%%%%%%%%%%%%%%%%%%%%%%%%%%%%%%%%%%%%%%%%%%%%
\par
In the thermal equilibrium FF state, when ${\bm Q}$ points to the $z$ direction, the pair amplitude has large intensity around the Fermi surface in the region $p_z>0$, as shown in Fig. \ref{fig_NFF_FF}(a). Thus, in Figs. \ref{fig_NESS}(b3) and (b4), the anisotropic pair amplitude in the upper-half (lower-half) plane is dominated by the first term $P_{\bm p}^{\rm FF}({\bm Q},\delta\mu)$ (second term ($P_{-{\bm p}}^{\rm FF}({\bm Q},-\delta\mu)$) in Eq. (\ref{eq.PA201}). On the other hand, the vanishing region (which is also referred to the blocking region in the superconductivity literature) spreads over the lower hemisphere in the thermal equilibrium FF state (see Fig. \ref{fig_NFF_FF}(a)). Noting that NFF may be viewed as a mixture of two FF states with ${\bm Q}$ and $-{\bm Q}$ as shown in Figs. \ref{fig_NFF_FF}(a) and \ref{fig_NFF_FF}(b), we find that their blocking regions give the vanishing pair amplitude around the equator in Fig. \ref{fig_NESS}(b3). 
\par
%%%%%%%%%%%%%%%%%%%%%%%%%%%%%%%%%%%%%%%%%%%%%%%%%%%%%%%%%%%%%%%%%%%%%%%%%%%%%%%%
\begin{figure}[tb]
\centering
\includegraphics[width=8cm]{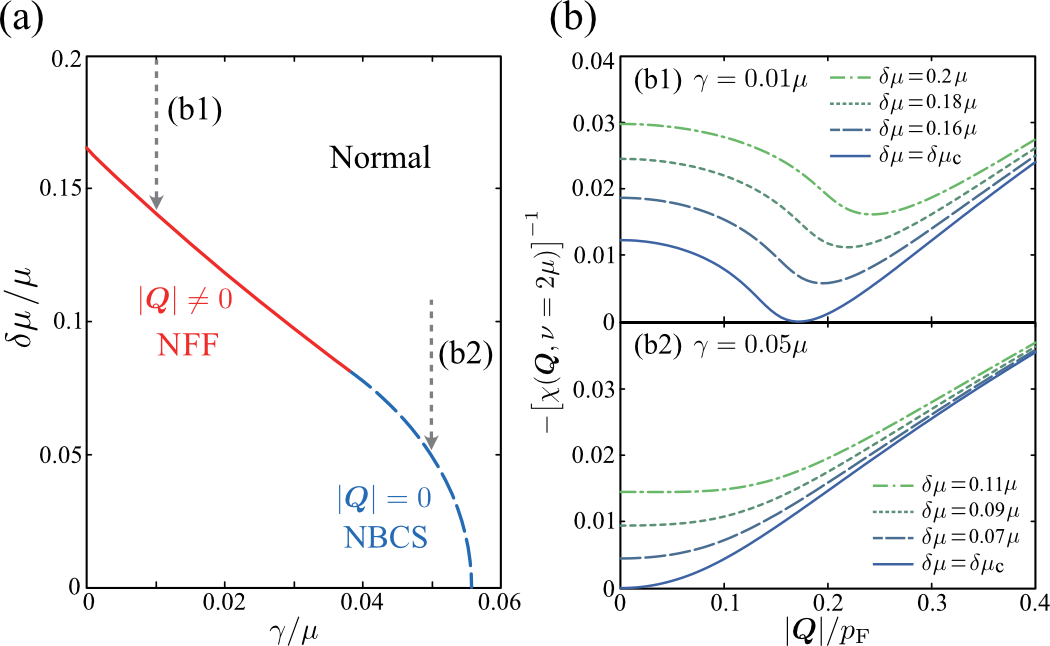}
\caption{
(a) Calculated $\delta\mu$ at the superfluid phase transition in the model driven-dissipative Fermi gas in Fig. \ref{fig1}(a). We take $T_{\rm env}=0$ and $(p_{\rm F}a_s)^{-1}=-1$. The system exhibits the NFF (NBCS) superfluid instability on the solid (dashed) line. (b) Inverse particle-particle scattering vertex $[\chi(\bm{Q}, \nu=2\mu) ]^{-1}$ in Eq. \eqref{eq.Thouless}, as a function of $\bm{Q}$. Upper (lower) panel shows the result along the path (b1) (path (b2)) in (a).}
\label{fig8} 
\end{figure}
%%%%%%%%%%%%%%%%%%%%%%%%%%%%%%%%%%%%%%%%%%%%%%%%%%%%%%%%%%%%%%%%%%%%%%%%%%%%%%%%
\par
Because the two-edge structure of the Fermi momentum distribution $n_{{\bm p},\sigma}$ at $p_{\rm F1}$ and $p_{\rm F2}$ are essentially important in obtaining NIG, NFF, and NFF' solutions, these states are expected to be suppressed when this structure is blurred with increasing the environment temperature $T_{\rm env}$. This can be confirmed in Fig. \ref{fig_finiteT}, where one sees that only the NBCS state remains at $T_{\rm env}=0.08\mu$. Of course, NBCS also eventually disappears at higher $T_{\rm env}$ as in the thermal equilibrium case, although we do not explicitly show the result here. 
\par
We note that, while the chemical potential difference $\delta\mu$ produces the two-edge structure $p_{\rm F1}$ and $p_{\rm F2}$ in $n_{{\bm p},\sigma}$, this structure may also be viewed as the smearing of the Fermi surface edge at $p_{\rm F}=\sqrt{2m\mu}$. Because this is a similar effect to the thermal broadening, all the four solutions eventually disappear when $\delta\mu$ is large to some extent, as shown in Figs. \ref{fig_NESS}(a) and \ref{fig_finiteT}. 
\par
We also note that the same depairing mechanism also works when the damping rate $\gamma$ becomes large. Figure \ref{fig8}(a) shows the superfluid phase transition line in the $\gamma$-$\delta\mu$ plane ($T_{\rm env}=0$), determined from the pole condition of the particle-particle scattering vertex $\chi(\bm{Q},\nu)$ in Eq. \eqref{eq.Thouless}. Because the damping $\gamma$ makes the two-step structure in $n_{{\bm p},\sigma}$ obscure \cite{Kawamura2020}, the superfluid instability of the NFF (${\bm Q}\ne 0$) changes to the BCS-type phase transition with ${\bm Q}=0$, when $\gamma/\mu\gesim 0.04$ (see Figs. \ref{fig8}(b1) and \ref{fig8}(b2)). As one further increases $\gamma$, the two steps in $n_{{\bm p},\sigma}$ are completely smeared out and the overall structure becomes similar to the thermal equilibrium case at high temperatures. As a result, the main system is in the normal state, when $\gamma/\mu\gesim 0.056$ in Fig. \ref{fig8}(a).
\par
\subsection{Stability analysis of non-equilibrium superfluid solutions}
\label{subsec_stability}
\par
We next assess the stability of the four non-equilibrium superfluid solutions (NBCS, NIG, NFF, and NFF') obtained in Sec. IV.A, by solving the linearized quantum kinetic equation (\ref{eq_linearized_QKE}) under the initial condition in Eq. \eqref{eq_initial}. 
Table I summarizes our result for  $T_{\rm env}=0$ and $\gamma=0.005\mu$. For finite bath temperature results, see also Fig. \ref{fig2}(a), which shows the same sets of the solution to be stable. We found that BCS and NFF are stable steady-state solutions in all regions, while NIG and NFF' are unstable.
\par
%%%%%%%%%%%%%%%%%%%%%%%%%%%%%%%%%%%%%%%%%%%%%%%%%%%%%%%%%%%%%%%%%%%%%%%%%%%%%%%%
\begin{figure}[tb]
\centering
\includegraphics[width=7.8cm]{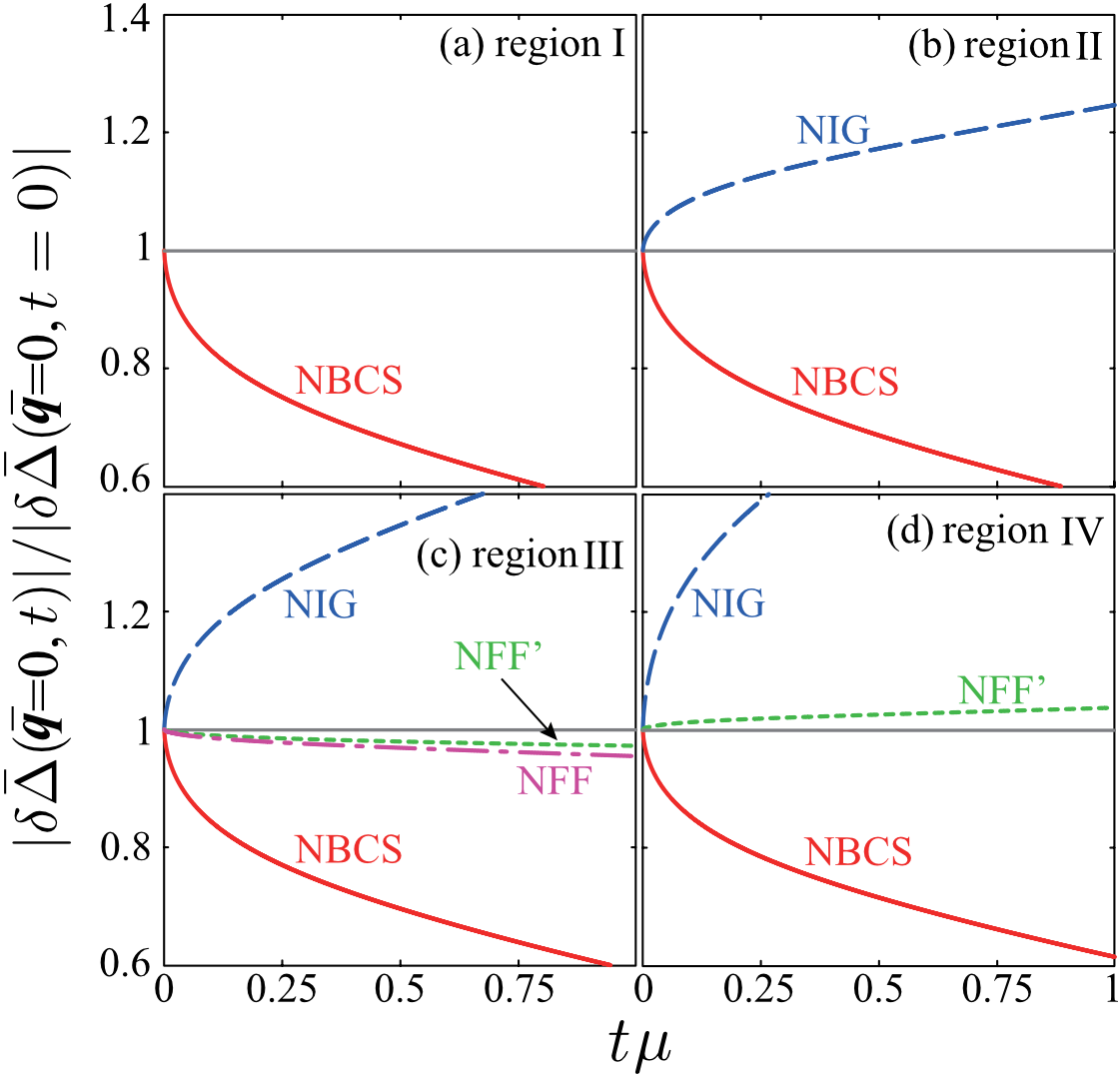}
\caption{Calculated time evolution of the deviation $|\delta\bar{\Delta}({\bar {\bm q}}=0,t)|$ of the superfluid order parameter from the mean-field value. We set $T_{\rm env}=0$, $\gamma=0.005\mu$, and $\delta|\bar{\Delta}({\bar {\bm q}}=0,t=0)|=0.001\mu$. (a) $\delta\mu=0.05\mu$ (region I). (b) $\delta\mu= 0.13\mu$ (region II). (c) $\delta\mu= 0.145\mu$ (region III). (d) $\delta\mu= 0.16\mu$ (region IV).}
\label{fig_stability1} 
\end{figure}
%%%%%%%%%%%%%%%%%%%%%%%%%%%%%%%%%%%%%%%%%%%%%%%%%%%%%%%%%%%%%%%%%%%%%%%%%%%%%%%%
\par
%%%%%%%%%%%%%%%%%%%%%%%%%%%%%%%%%%%%%%%%%%%%%%%%%%%%%%%%%%%%%%%%%%%%%%%%%%%%%%%%
\begin{figure}[tb]
\centering
\includegraphics[width=7.8cm]{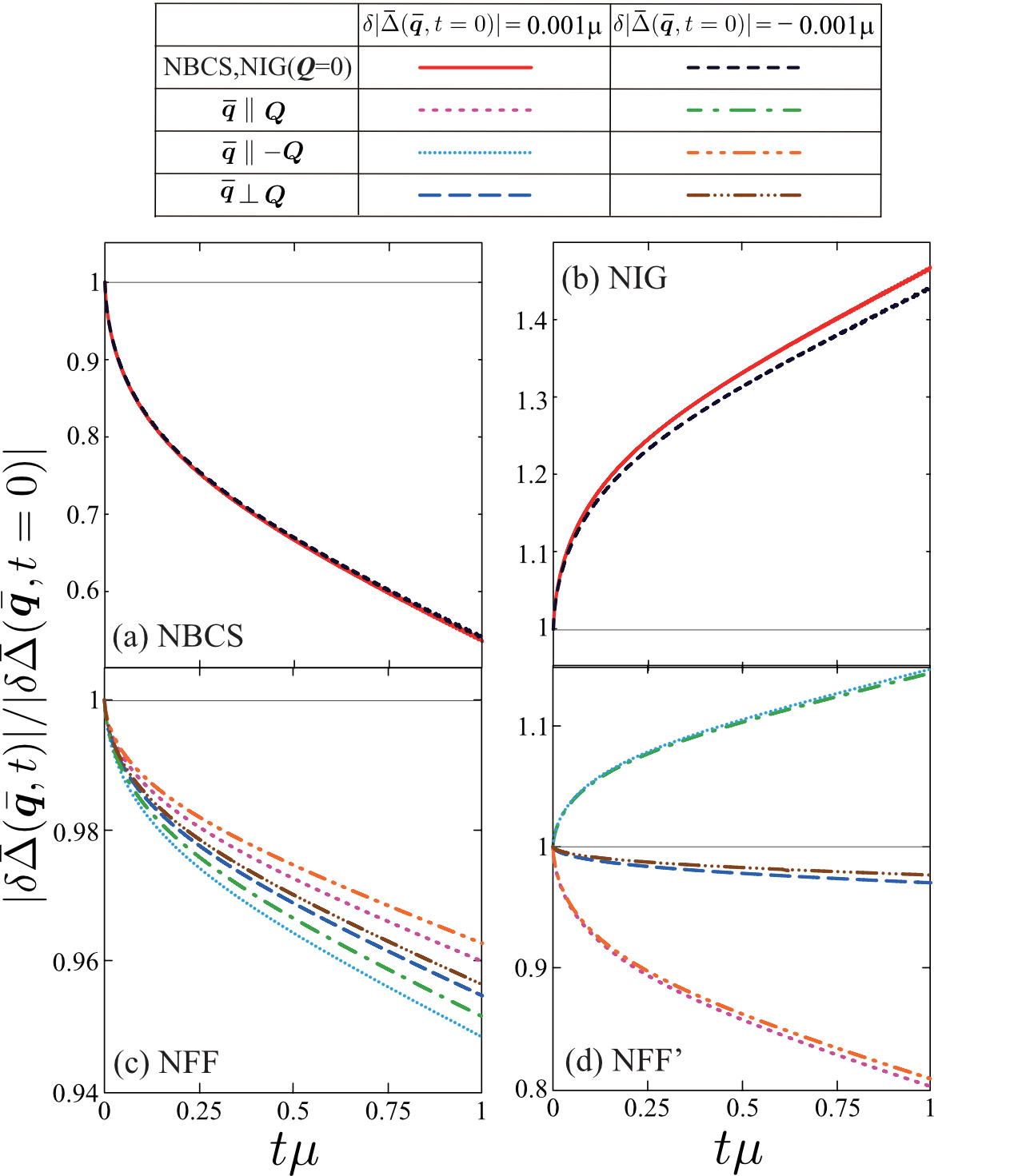}
\caption{Calculated time evolution of $|\delta\bar{\Delta}({\bar {\bm q}},t)|$ in the region III ($\delta\mu=0.145\mu$), when $|{\bar {\bm q}}|=0.001p_{\rm F}>0$. We take $T_{\rm env}=0$, $\gamma=0.005\mu$. (a) NBCS. (b) NIG. (c) NFF. (d) NFF'. In this figure, ``${\bm {\bar q}}\parallel\pm{\bm Q}$" show the cases when ${\bar {\bm q}}$ is parallel and points to $\pm{\bm Q}$. Because NBCS and NIG are isotropic with ${\bm Q}=0$, the results shown in panels (a) and (b) do not depend on the direction of ${\bar {\bm q}}$.}
\label{fig_stability2} 
\end{figure}
%%%%%%%%%%%%%%%%%%%%%%%%%%%%%%%%%%%%%%%%%%%%%%%%%%%%%%%%%%%%%%%%%%%%%%%%%%%%%%%%
\par
We report below the stability analysis result that confirms these results. Figure \ref{fig_stability1} shows the time evolution of the deviations of the superfluid order parameter from the mean-field value, when the amplitude deviation is only considered at $t=0$ (${\bar {\bm q}}=0$ and $\delta|\Delta({\bar {\bm q}}=0,t=0)|\ne 0$). In the NBCS case, we see in this figure that the deviation $|\delta\Delta({\bar {\bm q}}=0)|$ decays over time in all the regions I-IV.  This means that NBCS is stable against small perturbation in terms of the amplitude of the superfluid order parameter. The same conclusion is also obtained in the presence of phase deviation (${\bar {\bm q}}\ne 0$). As an example, we show in Fig. \ref{fig_stability2}(a) the result in the region III. Thus, we judge that NBCS is a {\it stable} non-equilibrium superfluid steady state.
\par
In contrast to NBCS, NIG exhibits the opposite behavior, as seen in Figs. \ref{fig_stability1}(b)-(d): The deviation $|\delta\Delta({\bar {\bm q}}=0,t)|$ grows over time, indicating that, although NIG is one of the four mean-field solutions, it is actually destroyed by this small perturbation. The instability of this state is also seen when ${\bar {\bm q}}\ne 0$, as shown in Fig. \ref{fig_stability2}(b). These results conclude that NIG is {\it unstable}.
\par
For the anisotropic solutions with ${\bm Q}\ne 0$, Figs. \ref{fig_stability1}(c) and \ref{fig_stability2}(c) conclude that NFF is a {\it stable} FF-type superfluid state in the region III. For NFF', as far as the perturbation with ${\bar {\bm q}}=0$ is considered, while it is unstable in the region IV, it is stable in the region III (see Figs. \ref{fig_stability1}(d) and (c), respectively). However, even in the latter region, Fig. \ref{fig_stability2}(d) shows that this FF-type state cannot be always stable against the initial deviation with ${\bar {\bm q}}\ne 0$. In this sense, we classify NFF' as an {\it unstable} superfluid state.
\par
%%%%%%%%%%%%%%%%%%%%%%%%%%%%%%%%%%%%%%%%%%%%%%%%%%%%%%%%%%%%%%%%%%%%%%%%%%%%%%%%
\begin{figure}[tb]
\centering
\includegraphics[width=7.8cm]{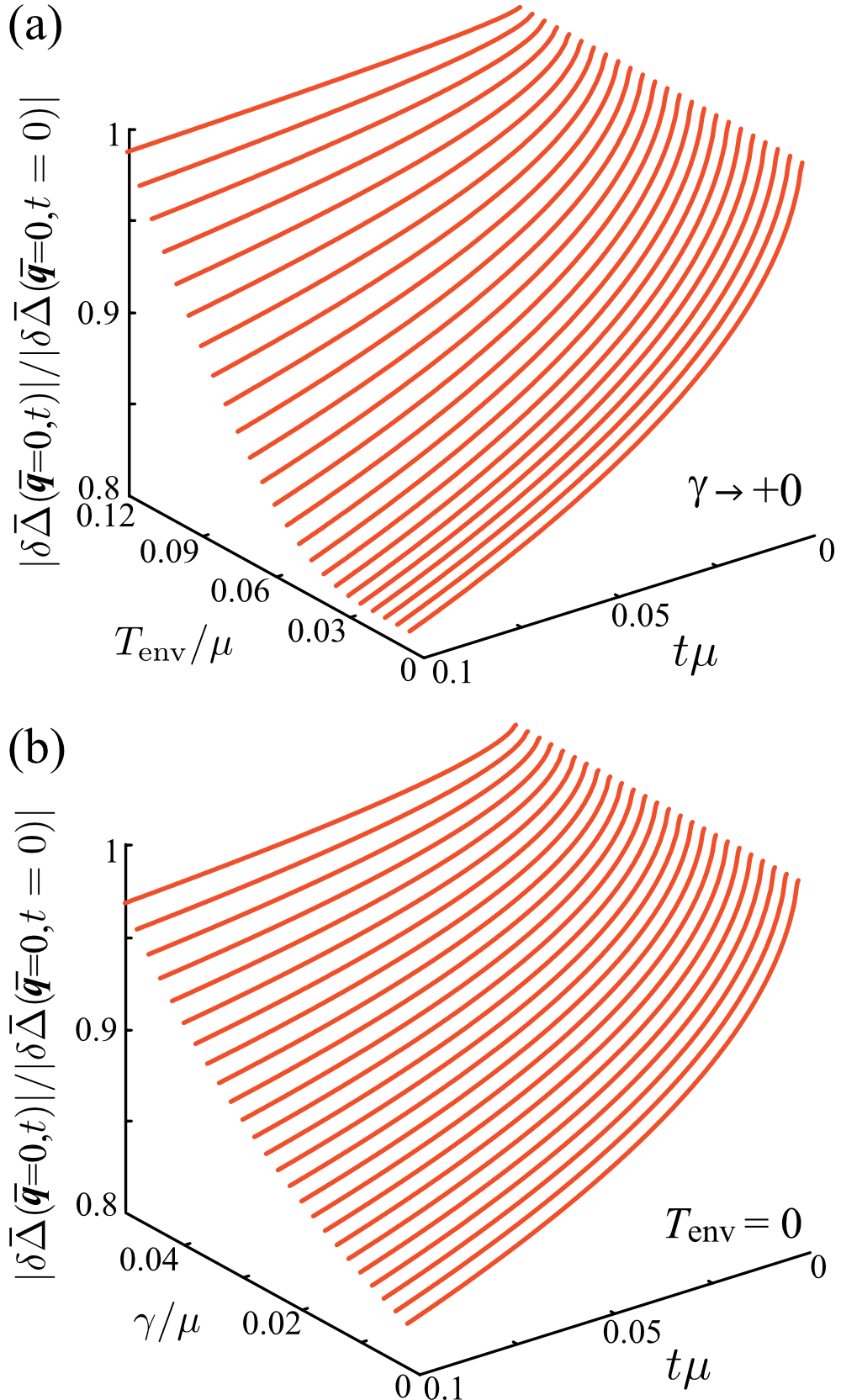}
\caption{Time evolution of the deviation $|\delta \bar{\Delta}({\bm q}=0,t)|$ from the NBCS mean-field order parameter and effects of (a) environment temperature $T_{\rm env}$, and (b) damping rate $\gamma$. We take $\delta\mu =0.05\mu$, and $\delta|\bar{\Delta}({\bm q}=0,t=0)|=0.001\mu$. In panels (a) and (b), we set $\gamma\to +0$ and $T_{\rm env}=0$, respectively.}
\label{fig_stability3}  
\end{figure}
%%%%%%%%%%%%%%%%%%%%%%%%%%%%%%%%%%%%%%%%%%%%%%%%%%%%%%%%%%%%%%%%%%%%%%%%%%%%%%%%
\par
The above conclusions hold true even in the case with finite bath temperature $T_{\rm env}>0$ and $\gamma>0$: We show in Fig. \ref{fig_stability3} effects of the environment temperature $T_{\rm env}$ (panel (a)) and damping rate $\gamma$ (panel (b)) on the time evolution of $|\delta\Delta({\bar {\bm q}}=0,t)|$ in the NBCS case: Initial deviations is found to always decay over time, irrespective of the values of these environment parameters.
\par
We also find from Fig. \ref{fig_stability3} that the relaxation time (time scale to recover the mean-field solution) depends on the environment parameters, $T_{\rm env}$ and $\gamma$. For the damping rate $\gamma$, we find from Fig. \ref{fig_stability3}(b) that it would be preferable to make the value of $\gamma/\mu$ as small as possible for realizing the NFF. If our proposed setup is implemented by using current experimental techniques in a cold atomic system, we expect that $\gamma$ can be reduced to about 0.01 $\mu$K and $\mu$ can be increased to about 1 $\mu$K, thereby keeping the value of $\gamma/\mu$ below 0.01, which is small enough to realize the NFF (see also Fig. \ref{fig2}(a)). How to estimate these values and more detailed discussions are given in Appendix C.
\par
Summarizing the above-mentioned stability analyses for various parameter sets ($T_{\rm env}, \gamma, \delta\mu$), we obtain the phase diagram in Figs. \ref{fig2}(a) and \ref{fig2}(b). Among the four candidates (NBCS, NIG, NFF, NFF'), NBCS and NFF only survive as stable non-equilibrium superfluid steady states. In the superfluid region, NBCS is always stable. Thus, in the region where NFF is stable, the bistability occurs. We also see in Figs. \ref{fig2}(a) and \ref{fig2}(b) the existence of the other bistability region (region(III)) where NBCS and the normal state are both stable. 
\par 
Because the energetic consideration, which is useful in determining the {\it thermodynamically stable state}, does not work in the present non-equilibrium case, one cannot immediately identify which state is realized in the bistability region. The answer to this question is considered to depend on how to tune the environment parameters: As $\delta\mu$ increases adiabatically from $\delta\mu=0$, the NBCS would be maintained both in the regions (II) and (III). As one decreases $\delta\mu$ from the region (IV), on the other hand, the phase transition from the normal state to NFF occurs at the boundary between (II) and (III) \cite{noteBoundary}. As a result, the superfluid order parameter $\Delta_0$ would exhibit the hysteresis behavior shown in Fig. \ref{fig2}(c). We note that Refs. \cite{Snyman2009, Bobkova2014, Ouassou2018} clarify that a voltage-driven superconductor (where the momentum distribution of conduction electrons is highly non-thermal and exhibits a two-step structure as in the non-equilibrium case we are considering in this paper) shows the same kind of bistability, although the possibility of FF-like state has not been discussed. Our result is consistent with these previous work, and NFF predicted in this paper would also be expected in such a voltage-driven superconductor.
\par
Finally, we comment on the experimental observational advantage of the NFF state compared to conventional FF-type superfluid/superconducting states in thermal equilibrium spin-imbalanced Fermi gases \cite{Hu2006, Parish2007, Liao2010, Chevy2010, Kinnunen2018, Strinati2018} and metallic superconductors under an external magnetic field \cite{Fulde1964, Larkin1964, Takada1969, Shimahara1994, Matsuda2007}. Although a spin-imbalanced Fermi gas is simpler than the setup proposed in this paper, which makes, at a glance, the former to be an ideal system for the realization of FF-type superfluid states, no clear observation of this exotic state has been reported so far. One reason for this difficulty is that the realization of an FF-type superfluid state in a spin-imbalanced Fermi gas always suffers from the phase separation (where the BCS state and the normal state coexist). Indeed, in previous experiments, only phase separation has been observed \cite{Partridge2006, Shin2006, Shin2008}. This happens because the experimental system is done with a fixed number of particles $N_\up$ and $N_\down$, while the chemical potential $\mu_\up$ and $\mu_\down$ are not fixed \cite{Bedaque2003, Caldas2004}. On the other hand, in our proposed setup, the main system would be controlled by the fixed chemical potential of the two reservoirs $\mu_{\rm L}$ and $\mu_{\rm R}$ while the number of particles is {\it not} fixed. Moreover, the NFF state realized by the Fermi-surface reservoir-engineering does {\it not} need any spin imbalance. Thus, our proposal can avoid the occurrence of the unwanted phase separation phenomenon, which is a clear experimental observational advantage compared to the conventional proposal in a thermal equilibrium cold atomic system.
\par
A metallic superconductors under an external magnetic field is also considered to be an ideal system for observing the FF state and has been vigorously studied. However, unambiguous experimental evidence for the pure FF state is still lacking in experiments. A major obstacle to observing the pure FF state in it is the orbital pair-breaking effect, which leads to a mixing of the FF state and the Abrikosov vortex state \cite{Shimahara1997, Shimahara2009}. However, using our proposed Fermi-surface reservoir-engineering instead of an external magnetic field, one can clearly avoid this problem: As discussed in Sec. \ref{subsec:model}, a voltage-biased superconducting wire or thin film is also a promising experimental setup to observe the NFF state. Since there is no external field coupled to the spatial motion of electrons in these setups, in principle, we can realize a pure FF-type superconducting state in any (clean) superconducting metals by using our proposed engineering scheme.
\par
Thus, while the previous thermal equilibrium approaches do not succeed in observing clear FF-type superfluid/superconducting state, we expect that the Fermi-surface reservoir-engineering is a promising alternative route to reach this inhomogeneous pairing state.
%%%%%%%%%%%%%%%%%%%%%%%%%%%%%%%%%%%%%%%%%%%%%%%%%%%%%%%%%%%%%%%%%%%%%%%%%%%%%%%%
\par
\section{Summary}
\par
To summarize, we have proposed an idea to process the Fermi momentum distribution $n_{{\bm p},\sigma}$, by using reservoirs with different chemical potentials. Although we expect our scheme to work for generic Fermi systems, as a paradigmatic example, we have discussed non-equilibrium superfluid steady states and their stability in a driven-dissipative two-component Fermi gas. Using the Nambu-Keldysh Green's function technique, we extended the BCS theory developed in the thermal equilibrium state to the non-equilibrium steady state. To examine the stability of steady-state solutions obtained from this non-equilibrium BCS scheme, we also derived a quantum kinetic equation, to examine the time evolution of the superfluid order parameter. 
\par
By solving the non-equilibrium gap equation, we obtained four superfluid steady-state solutions: Among them, one of them (NBCS state) may be viewed as an extension of the ordinary thermal equilibrium BCS state to the non-equilibrium case, and another one (NIG state) is similar to the Sarma(-Liu-Wilczek) interior gap state. While these are isotropic uniform states, the remaining two are Fulde-Ferrell (FF) like anisotropic and non-uniform superfluid states (NFF and NFF' state), even though the present system has no spin imbalance. Analyzing their pair amplitudes, we found that the latter three superfluid solutions originate from the two-edge structure of the non-equilibrium Fermi momentum distribution which is produced by the coupled two reservoirs with different chemical potentials. We also pointed out that each FF-like state may be viewed as the superposition of the thermal equilibrium FF state under an external magnetic field $h=\delta\mu$ and that under an external magnetic field $h=-\delta\mu$.
\par
We then studied the stability of these four superfluid steady-state solutions, by solving the linearized quantum kinetic equation. This concluded that only the BCS-type state (NBCS) and one of the two FF-type states (NFF) are stable, in the sense that the initial deviation of the superfluid order parameter always decays over time. The other two, NIG and NFF', are unstable because small perturbation are amplified over time. These stability analyses lead to the phase diagram of a driven-dissipative Fermi gas shown in Figs. \ref{fig2}(a) and \ref{fig2}(b). 
\par
Our proposed Fermi-surface reservoir-engineering can be applied not only to the Fermi gas system but also to various many-body Fermi systems. Particularly in lattice systems, the combination of the band structure and multi-step structure on the Fermi momentum distribution may trigger unconventional ordered phases, such as spin- and charge-density wave-like states. The stability of such unconventional ordered phases can be assessed by evaluating the time evolution of fluctuations around steady-state value, in the same manner as this paper. The search for unconventional ordered phases in non-equilibrium systems is currently one of the most exciting challenges in the condensed matter physics, and our results would contribute to the further development of this research field.
\par
%%%%%%%%%%%%%%%%%%%%%%%%%%%%%%%%%%%%%%%%%%%%%%%%%%%%%%%%%%%%%%%%%%%%%%%%%%
\par
\begin{acknowledgments}
\par
We thank D. Kagamihara, K. Furutani, and M. Horikoshi for discussions. T.K. was supported by MEXT and JSPS KAKENHI Grant-in-Aid for JSPS fellows Grant No.JP21J22452. R.H. was supported by an appointment to the JRG Program at the APCTP through the Science and Technology Promotion Fund and Lottery Fund of the Korean Government. Y.O. was supported by a Grant-in-aid for Scientific Research from MEXT and JSPS in Japan (No.JP18K11345, No.JP18H05406, and No.JP19K03689).
\end{acknowledgments}
\par
%%%%%%%%%%%%%%%%%%%%%%%%%%%%%%%%%%%%%%%%%%%%%%%%%%%%%%%%%%%%%%%%%%%%%%%%%%%%%%%%
\par
\appendix
\par
\section{Derivation of Eq. (\ref{EOS_Glesser})}
\par
We first introduce two inverse Green's functions $\overrightarrow{\m{G}}^{-1}_0(1)$ and $\overleftarrow{\m{G}}^{-1}_0(2)$, that obey
\begin{align}
& \overrightarrow{\m{G}}^{-1}_0(1) \m{G}^{\rm R(A)}_0(1,2) = \delta(1-2) \tau_0, \\[4pt]
& \m{G}^{\rm R(A)}_0(1,2) \overleftarrow{\m{G}}^{-1}_0(2) = \delta(1-2) \tau_0,
\label{eq.A1}
\end{align}
where $\m{G}^{\rm R(A)}_0(1,2)$ is the retarded (R) or advanced (A) component of the bare Green's function in Eq. (\ref{eq.5}), $\delta(1-2)=\delta({\bm r}_1-{\bm r}_2)\delta(t_1-t_2)$, and the left (right) arrow on each differential operator means that it acts on the left (right) side of this operator. From the Heisenberg equation of the field operator, these inverse Green's functions are found to have the forms,
\begin{align}
& \overrightarrow{\m{G}}^{-1}_0(1) =
\left( \begin{array}{cc}
i\overrightarrow{\partial}_{t_1} -h_0(-i\overrightarrow{\nabla}_{\bm{r}_1}) & 0 \\
0 & i\overrightarrow{\partial}_{t_1} -h_0(-i\overrightarrow{\nabla}_{\bm{r}_1})
\end{array} \right), \\[4pt]
& \overleftarrow{\m{G}}^{-1}_0(2) =
\left( \begin{array}{cc}
-i\overleftarrow{\partial}_{t_2} -h_0(i \overleftarrow{\nabla}_{\bm{r}_2}) & 0 \\
0 & -i\overleftarrow{\partial}_{t_2} -h_0(i\overleftarrow{\nabla}_{\bm{r}_2})
\end{array} \right),
\end{align}
where $h_0(-i\nabla_{\bm r})=(-i\nabla_{\bm r})^2/(2m)$. Carrying out the gauge transformation discussed in Eqs. (\ref{eq.gaugeA})-(\ref{eq.gaugeC}), one has
\begin{widetext}
\begin{align}
\overrightarrow{\tilde{\m{G}}}^{-1}_0(1) 
&\equiv
e^{-i\chi(1) \tau_3}\overrightarrow{\m{G}}^{-1}_0(1) e^{i\chi(1) \tau_3}
=
\left( 
\begin{array}{cc}
i\overrightarrow{\partial}_{t_1} -h_0(-i\overrightarrow{\nabla}_{\bm{r}_1} +\bm{Q}/2) +\mu& 0 \\
0 & i\overrightarrow{\partial}_{t_1} +h_0(-i\overrightarrow{\nabla}_{\bm{r}_1} +\bm{Q}/2) +\mu
\end{array} 
\right)
\label{eq_Ginv1}
,\\[8pt]
\overleftarrow{\tilde{\m{G}}}^{-1}_0(2) 
&\equiv
e^{-i\chi(2) \tau_3} \overleftarrow{\m{G}}^{-1}_0(2) e^{-i\chi(2) \tau_3}
=
\left( 
\begin{array}{cc}
-i\overleftarrow{\partial}_{t_2} -h_0(i \overleftarrow{\nabla}_{\bm{r}_2} +\bm{Q}/2) +\mu& 0 \\
0 & -i\overleftarrow{\partial}_{t_2} +h_0(i\overleftarrow{\nabla}_{\bm{r}_2} +\bm{Q}/2)+\mu
\end{array} 
\right).
\label{eq_Ginv2}
\end{align}
\par
Operating $\overrightarrow{\tilde{\m{G}}}^{-1}_0(1)$ and $\overleftarrow{\tilde{\m{G}}}^{-1}_0(2)$ to the Dyson equation of the (gauge transformed) lesser Green's function \cite{Rammer2007,Zagoskin2014},
\begin{equation}
\tilde{\m{G}}^<(1,2)=\big[\tilde{\m{G}}^{\rm R}\circ\tilde{\Sigma}^<\circ\tilde{\m{G}}^{\rm A} \big](1,2),
\label{GL_Dyson_tilde}
\end{equation}
from the left and the right, we have, respectively,
\begin{align}
& \overrightarrow{\tilde{\m{G}}}^{-1}_0(1) \tilde{\m{G}}^{<}(1,2)
=\big[\tilde{\Sigma}^<\circ \tilde{\m{G}}^{\rm A} 
+ \tilde{\Sigma}^{\rm R} 
\circ \tilde{\m{G}}^< \big](1,2)
\label{eq_LDyson_L},
\\
& \tilde{\m{G}}^{<}(1,2) \overleftarrow{\tilde{\m{G}}}_0^{-1}(2)
=\big[\tilde{\m{G}}^{\rm R} \circ \tilde{\Sigma}^< 
+  \tilde{\m{G}}^< \circ \tilde{\Sigma}^{\rm A}\big](1,2).
\label{eq_RDyson_L}
\end{align}
In obtaining these equations, we have used
\begin{align}
& \overrightarrow{\tilde{\m{G}}}^{-1}_0(1) 
\tilde{\m{G}}^{\rm R(A)}(1,2) 
= \delta(1-2) \tau_0 
+ \big[\tilde{\Sigma}^{\rm R(A)} \circ \tilde{\m{G}}^{\rm R(A)} \big](1,2)
\label{eq_LDyson_L2},
\\
& \tilde{\m{G}}^{\rm R(A)}(1,2) 
\overleftarrow{\tilde{\m{G}}}_0^{-1}(2) 
= \delta(1-2) \tau_0 
+ \big[ \tilde{\m{G}}^{\rm R(A)} \circ \tilde{\Sigma}^{\rm R(A)} \big](1,2). 
\label{eq_RDyson_L2}
\end{align}
Equations \eqref{eq_LDyson_L} and \eqref{eq_RDyson_L} yield the Kadanoff-Baym (KB) equation \cite{Baym1961, Rammer2007}, 
\begin{equation}
\big[
\overrightarrow{\tilde{\m{G}}}^{-1}_0  \tilde{\m{G}}^{<} 
-
\tilde{\m{G}}^{<} \overleftarrow{\tilde{\m{G}}}^{-1}_0 
\big](1,2)
=\big[ 
\tilde{\Sigma}^{\rm R} \circ \tilde{\m{G}}^< 
-\tilde{\m{G}}^< \circ \tilde{\Sigma}^{\rm A} 
+\tilde{\Sigma}^<\circ \tilde{\m{G}}^{\rm A} 
-\tilde{\m{G}}^{\rm R} \circ \tilde{\Sigma}^<  
\big](1,2) 
\label{eq_KB}.
\end{equation}
Carrying out the Wigner transformation, which is followed by the 
$\omega$-integration, one finds that the KB equation (\ref{eq_KB}) becomes
\begin{align}
\overrightarrow{\tilde{\m{G}}}^{-1}_0 
\bar{\m{G}}^{<}_{\bm{p}}(\bm{r},t) 
-\tilde{\m{G}}^<_{\bm{p}}(\bm{r},t) 
\overleftarrow{\tilde{\m{G}}}^{-1}_0
= \m{I}_{\bm{p}}(\bm{r},t), 
\label{eq_KB2} 
\end{align}
where $\m{I}_{\bm{p}}(\bm{r},t)$ is given in the sum of Eqs. (\ref{eq.collA}) and (\ref{eq.collB}). Since the left hand side of Eq. (\ref{eq_KB2}) is evaluated as
\begin{eqnarray}
\overrightarrow{\tilde{\m{G}}}^{-1}_0 \tilde{\m{G}}^{<}_{\bm{p}}(\bm{r},t) 
&-&
\tilde{\m{G}}^<_{\bm{p}}(\bm{r},t)\overleftarrow{\tilde{\m{G}}}^{-1}_0 
=
i\partial_t \tilde{\m{G}}^<_{\bm{p}}(\bm{r},t) -
\big[\xi_{\bm{p}}\tau_3, \tilde{\m{G}}^<_{\bm{p}}(\bm{r},t) \big]_- 
-
\left[\frac{{\bm Q}^2}{8m}\tau_3,  \tilde{\m{G}}^<_{\bm{p}}(\bm{r},t) \right]_- 
\nonumber
\\[4pt]
&{ }&
\hskip-20mm
+\left[\frac{1}{8m}\tau_3, \nabla_{\bm r}^2 \tilde{\m{G}}^<_{\bm{p}}(\bm{r}, t)\right]_-
+\frac{i}{2}\left[{\bm{p} \over m}\tau_3, \nabla_{\bm r} \tilde{\m{G}}^<_{\bm{p}}(\bm{r}, t)\right]_+
+\frac{i}{2} \left[ \frac{\bm{Q}}{2m} \tau_0, \nabla_{\bm r} \tilde{\m{G}}^<_{\bm{p}}(\bm{r}, t)\right]_+,
\end{eqnarray}
one reaches Eq. (\ref{EOS_Glesser}).
\par
%%%%%%%%%%%%%%%%%%%%%%%%%%%%%%%%%%%%%%%%%%%%%%%%%%%%%%%%%%%%%%%%%%%%%%%%%%%%%%%
\par
\section{Vanishment of the last term in Eq. (\ref{eq_Ienv_Wig})}
\par
To evaluate ${\m R}({\bm p},\omega,{\bm r},t)$ in Eq. (\ref{eq.R}), we carry out the Wigner transformation of Eqs. (\ref{eq_LDyson_L2}) and (\ref{eq_RDyson_L2}). Then, retaining the Moyal product to the first-order gradient expansion explained below Eq. (\ref{eq_Moyal}), we have
\begin{align}
\overrightarrow{\tilde{\m{G}}}^{-1}_0 \tilde{\m{G}}^{\rm R(A)}(\bm{p},\omega, \bm{r},t) 
&=
\tau_0 + \big[\tilde{\Sigma}^{\rm R(A)} \circ \tilde{\m{G}}^{\rm R(A)} \big](\bm{p},\omega, \bm{r}, t) 
\simeq
\tau_0 + \tilde{\Sigma}^{\rm R(A)}(\bm{p},\omega, \bm{r},t) \tilde{\m{G}}^{\rm R(A)}(\bm{p},\omega, \bm{r},t)
\label{eq_LDyson_R_Wigner}
,\\[6pt]
\tilde{\m{G}}^{\rm R(A)}\overleftarrow{\tilde{\m{G}}}_0^{-1}(\bm{p},\omega, \bm{r},t)
&=
\tau_0 + \big[ \tilde{\m{G}}^{\rm R(A)} \circ \tilde{\Sigma}^{\rm R(A)} \big](\bm{p},\omega, \bm{r}, t) 
\simeq
\tau_0  +\tilde{\m{G}}^{\rm R(A)}(\bm{p},\omega, \bm{r},t) \tilde{\Sigma}^{\rm R(A)}(\bm{p},\omega, \bm{r},t).
\label{eq_RDyson_R_Wigner}
\end{align}
The sum of these equations gives
\begin{equation}
\left[
\omega \tau_0 -\xi_{\bm{p}} \tau_3 -\frac{{\bm Q}^2}{8m}\tau_3 
-\frac{1}{2} {{\bm p} \over m} \cdot \bm{Q} \tau_0 -\tilde{\Sigma}^{\rm R(A)}(\bm{p},\omega, \bm{r},t), 
\tilde{\m{G}}^{\rm R(A)}(\bm{p},\omega, \bm{r},t) 
\right]_+ = 2\tau_0.
\end{equation}
Here, ${\tilde \Sigma}^{\rm R(A)}({\bm p},\omega,{\bm r},t)={\tilde \Sigma}_{\rm int}^{\rm R(A)}({\bm p},\omega,{\bm r},t)+{\tilde \Sigma}_{\rm env}^{\rm R(A)}({\bm p},\omega,{\bm r},t)$. This equation yields
\begin{align}
&
\tilde{\m{G}}^{\rm R}(\bm{p}, \omega, \bm{r}, t) =
\sum_{\eta=\pm}
\frac{1}{\omega + 2i\gamma -E^\eta_{\bm{p},\bm{Q}}(\bm{r},t) } \Xi^\eta_{\bm{p},\bm{Q}}(\bm{r},t)
\label{eq_app_GR}
,\\[4pt]
&
\tilde{\m{G}}^{\rm A}(\bm{p}, \omega, \bm{r}, t) =
\sum_{\eta=\pm}
\frac{1}{\omega - 2i\gamma -E^\eta_{\bm{p},\bm{Q}}(\bm{r},t) } \Xi^\eta_{\bm{p},\bm{Q}}(\bm{r},t),
\label{eq_app_GA}
\end{align}
where
\begin{align}
& E^{\pm}_{\bm{p},\bm{Q}}(\bm{r},t) 
=\sqrt{\bigl( \xi^{\rm (s)}_{\bm{p},\bm{Q}}\bigr)^2 + |\bar{\Delta}(\bm{r},t)|^2} 
\pm \xi^{\rm a}_{\bm{p},\bm{Q}} 
\equiv E_{\bm{p}, \bm{Q}}(\bm{r},t) 
\pm \xi^{\rm (a)}_{\bm{p},\bm{Q}},
\label{app_E}
\\[6pt]
& 
\Xi^\pm_{\bm{p},\bm{Q}} = {1 \over 2}
\left[
\tau_0
\pm{\xi_{{\bm p},{\bm Q}}^{\rm (s)} \over E_{\bm{p}, \bm{Q}}(\bm{r},t)}\tau_3
\mp{{\tilde \Delta}({\bm r},t) \over E_{\bm{p}, \bm{Q}}(\bm{r},t)}
\right].
\label{app_M}
\end{align}
Equation (\ref{eq_app_GR}) may be viewed an extension of the Green's function in the non-equilibrium steady state in Eq. \eqref{eq_NESS_GR} to the case when the superfluid order parameter ${\bar \Delta}(\bm{r},t)$ depends on $t$ and ${\bm r}$ as given in Eq. (\ref{eq_N_OP}). 
\par
Subtracting Eqs. \eqref{eq_LDyson_R_Wigner} from \eqref{eq_RDyson_R_Wigner}, one obtains
\begin{equation}
i\partial_t \tilde{\m{G}}^{\rm R(A)}(\bm{p},\omega, \bm{r},t)  = \left[ \xi_{\bm{p}} \tau_3 + \frac{{\bm Q}^2}{8m} \tau_3 + \tilde{\Sigma}^{\rm R(A)}(\bm{p}, \omega, \bm{r}, t), \tilde{\m{G}}^{\rm R(A)}(\bm{p},\omega, \bm{r},t) \right]. 
\label{eq_Dyson_GR_sub}
\end{equation}
Because $\tilde{\m{G}}^{\rm R}(\bm{p},\omega, \bm{r},t)$ and $\tilde{\m{G}}^{\rm A}(\bm{p},\omega, \bm{r},t)$ are, respectively, given by Eqs. (\ref{eq_app_GR}) and (\ref{eq_app_GA}), the right hand side of Eq. (\ref{eq_Dyson_GR_sub}) is found to vanish, which immediately proves $\partial_t{\m R}({\bm p},\omega,{\bm r},t)=0$ in the last term in Eq. (\ref{eq_Ienv_Wig}).
\par
\end{widetext}
%%%%%%%%%%%%%%%%%%%%%%%%%%%%%%%%%%%%%%%%%%%%%%%%%%%%%%%%%%%%%%%%%%%%%%%%%%%%%%%
\par
\section{Detailed proposals for experiments}
\label{app:experiments}
In Sec. \ref{subsec:model}, we gave concrete proposals for experiments to realize our predicted non-equilibrium FF-type superfluid state (NFF). In order to achieve the NFF state, however, we foresee two experimental challenges. The first challenge is to achieve the parameters which can actually realize NFF. In particular, in addition to cooling down the left and right reservoirs to low enough temperature $T_{\rm L,R}\ll T_{\rm F}$ (where $T_{\rm F}=[3\pi^2 (n_\uparrow+n_\downarrow)]^{2/3}$ is the Fermi temperature of the main system), it is necessary for the linewidth $\gamma$ arising from the reservoir-system coupling to be kept below  $0.01\mu$ (See Figs. \ref{fig2}(a) and \ref{fig2}(b)). The second challenge (for ultracold atomic systems) is to maintain the system in the timescale of relaxation to the steady state. Since realistically, both the main system and the reservoirs are isolated and finite, the chemical potential bias between the reservoirs initially present, would eventually relax to its chemical equilibrium $\mu_{\rm L}=\mu_{\rm R}$. It is, therefore, important for the system to achieve the (quasi-) steady state much faster than the change of the chemical potential.
\par
In the following, we argue that current state-of-art experimental techniques of ultracold Fermi systems enable us to overcome these challenges. We first argue that it is possible to achieve $\gamma/\mu$ around 0.01 in our proposed cold atomic setup. Then, we explain that the relaxation to the steady state is expected to occur on a timescale sufficiently fast compared to the relaxation timescale of the chemical potential bias to allow us to observe the NFF.
\par
First, to address the first challenge, let us estimate the magnitude of the parameter $\gamma$. In the experiment of cold atomic transport measurements in a two-terminal configuration \cite{Brantut2012, Krinner2015, Husmann2015, Krinner2017, Husmann_thesis, Kanasz-Nagy2016}, quantized conductance is observed by sweeping the gate potential (corresponding to the gate voltage) at the quantum point contact (QPC) between the reservoirs. In the idealized situation with no broadening effects, the conductance is expected to change discontinuously from one plateau to the next when the gate potential is swept over a (transverse-)energy mode in the QPC. Realistically, however, the line broadening with the magnitude of about $0.01\mu$K is observed \cite{Krinner2015, Kanasz-Nagy2016}. This step broadening is considered to be mainly attributed to the following two factors; (i) finite temperature effects and (ii) broadening of the energy modes in the QPC due to the coupling with the reservoirs. The parameter $\gamma$ in our model is related to the latter. However, it is known that the broadening of the plateau is dominated by the finite temperature effect (i) \cite{Krinner2015, Krinner2017, Husmann_thesis} (where it shows good agreement with the experimental results even though the theoretical prediction in \cite{Krinner2015} neglects the influence of (ii)). Thus, we can reasonably estimate that $\gamma$ can be less than 0.01$\mu$K
\par
Next, we estimate the value of $\mu$. In order to make the model parameter $\gamma/\mu$ small, it is preferable to make the chemical potential $\mu$ as large as possible. Experimentally, we expect $\mu$ to be able to increase to about 1$\mu$K. In the recent transport experiments on a $^6$Li Fermi gas in a two-terminal configuration, the Fermi temperature of the atomic cloud is set from 300nK to 1$\mu$K \cite{Krinner2017}. Thus, the Fermi temperature of the main system, which is comparable to $\mu$, is also estimated to be possible to set to about 1$\mu$K. (Note that the Fermi temperature is determined by the particle density, and the density of the reservoirs and the main systems are comparable in our model.)
\par
From the above-mentioned analyses, we reasonably expect that it is possible to reduce the model parameter $\gamma/\mu$ to less than 0.01 when our proposed experimental setup is implemented in cold atomic systems. This enables the non-equilibrium distribution to show a clear two-step structure, which is a necessary ingredient to realize a stable NFF.
\par
We next discuss the second challenge for the implementation of our proposal: the timescale of the chemical potential bias decay and the relaxation to a steady state. In a typical cold atomic transport measurement, it is known that the relaxation of the chemical potential bias occurs on a time scale of a few seconds (for example, about 8s in \cite{Krinner2015}). On the other hand, the timescale of the relaxation to a steady state is considered to be dominated by interatomic scattering when the quasiparticle damping due to the system-reservoir couplings is sufficiently small ($\gamma \ll \mu$). The thermalization of incident particles from the QPC to the reservoir occurs on the scale of the interatomic scattering time $\tau=(1/n \sigma v_{\rm F})(T/T_{\rm F})\sim 400$ms, where $n$ is the particle density, $\sigma$ is the scattering cross-section for interparticle collisions, and $v_{\rm F}$ is the Fermi velocity \cite{Krinner2015}. As a result, our proposed setup is also expected to relax to a steady state in a few hundred milliseconds due to interatomic scattering. We note that the larger $n$ and $v_{\rm F}$ are, the shorter $\tau$ becomes and the more quickly the steady state can be achieved. This is compatible with the requirement that $\mu$ should be as large as possible in order to keep $\gamma/\mu$ small.
\par
To summarize, the relaxation to the non-equilibrium steady state we are interested in is expected to occur on a fast enough timescale ($\sim$ a few hundred milliseconds) compared to the timescale of the chemical potential bias decay ($\sim$ a few seconds), which indicates that it is quite promising to observe the NFF in our proposed cold atomic setup.
\par
So far, we have discussed the cold atomic setup, but voltage-biased superconductors would also satisfy the necessary condition for observing the NFF. Since a non-equilibrium quasiparticle distribution with a clear two-step structure has been experimentally observed in a voltage-biased electron system \cite{Pothier1997, Poither1997_2, Anthore2003, Chen2009}, we can reasonably expect that it is possible to make $\gamma/\mu$ small enough to realize the NFF. Actually, for the electron system, the damping due to impurity scattering would become more essential for the realization of NFF, rather than the damping $\gamma$ due to the system-reservoir couplings. Since impurities also have similar effects as $\gamma$ in our model, the two-step structure of the nonequilibrium distribution is smeared, which is detrimental to NFF, as the impurity concentration increases (Indeed, it has been observed that the two-step structure of a nonequilibrium distribution becomes blurred as the impurity concentration increases in mesoscopic wires \cite{Anthore2003}). This suggests that it is preferable to use clean (ballistic) samples in order to realize the NFF in voltage-biased electron systems.
%%%%%%%%%%%%%%%%%%%%%%%%%%%%%%%%%%%%%%%%%%%%%%%%%%%%%%%%%%%%%%%%%%%%%%%%%%%%%%
\par


\begin{thebibliography}{99}
\bibitem{noteT} We note that the temperature cannot be defined in the main system, when it is out of equilibrium. In the thermal equilibrium limit, the system temperature coincides with the environment temperature $T_{\rm env}$. In this paper, the term `temperature' also means $T_{\rm env}$ in the thermally equilibrium reservoirs.
\bibitem{noteP}  
Note that in our model, the number of atoms in the main system is not fixed, and the chemical potential difference between the reservoirs (=$\delta\mu$) is a natural tunable parameter. This situation is quite different from that of a thermal equilibrium spin-imbalanced Fermi gas, where the number of atoms is fixed and the number difference between the spin-up and -down component is a tunable parameter.
\bibitem{note} We note that the parameter $\mu$ in Eq. (\ref{eq_OP}) is given by the filling of the reservoirs, see Fig. \ref{fig1}(b).
\bibitem{noteBloch} Although, in equilibrium, the vanishing-current condition $\bm{J}_{\rm net}=0$ must always be satisfied in any thermodynamically stable state (which is sometimes referred to the Bloch's theorem in the literature \cite{Bloch,OhashiMomoi}), this theorem does not necessarily apply out of equilibrium.
\bibitem{note2} Although the last term in Eq. (\ref{eq_kinetic}) may affect the dynamics of quantities obtained without taking the $\bm{p}$-summation, such as the pair amplitude given by Eq. (\ref{eq.PA_def}), the following discussion will not handle the dynamics of such quantities.
\bibitem{note3} The Moyal product in Eq. \eqref{eq_Moyal} involves the product of $\bm{p}$-derivative and $\bm{r}$-derivative, such as $\overleftarrow{\partial_{\bm{r}}} \cdot \overrightarrow{\partial_{\bm{p}}}$. While the $\bm{r}$-derivative is on the order of $|\bar{\bm{q}}|$, the $\bm{p}$-derivative is estimated to be $p_{\rm F}^{-1}$: Since $\Delta_0 \ll \ep_{\rm F}$ (quasiclassical condition) is satisfied in the BCS regime, the Green's function is localized at $|\bm{p}|=p_{\rm F}$ in the momentum space. In this situation, we can fix the magnitude of the momentum at $p_{\rm F}$, that is, $\bm{p}=p_{\rm F} \hat{\bm{e}}_r +p_\theta \hat{\bm{e}}_\theta +p_\phi \hat{\bm{e}}_\phi$. Then, the gradient in the momentum space reads $\partial_{\bm{p}} = p_{\rm F}^{-1}\partial_{p_\theta}\hat{\bm{e}}_\theta +(p_{\rm F}\sin\theta)^{-1} \partial_{p_\phi} \hat{\bm{e}}_\phi$, and one can see that the $\bm{p}$-derivative is on the order of $p_{\rm F}^{-1}$ (see \cite{KopninBook} for more general and longer discussion). Thus, the terms arising from $\overleftarrow{\partial_{\bm{r}}} \cdot \overrightarrow{\partial_{\bm{p}}}$ are of order  $|\bar{\bm{q}}|/p_{\rm F}$, and the first-order gradient approximation explained below Eq. \eqref{eq_Moyal} will hold well when $|\bar{\bm{q}}|/p_{\rm F} \ll 1$ is satisfied.
\bibitem{noteGamma} Note that $\gamma$ is not exactly set to zero. In this paper, we use the notation $\gamma \to +0$ to mean $\gamma\to 0 +\delta$, where $\delta$ is an infinitesimally small positive number. In this limit, the main system is very weakly connected to reservoirs, just like the usual situation considered in the grand canonical ensemble. Then, although there is no damping due to the couplings to the reservoirs, the atomic distribution of the main system still reflects the distribution of the reservoirs.
\bibitem{noteBoundary} Note that the boundary between the region (II) and (II) in Fig. \ref{fig2} (b) corresponds to the solid line in Fig. \ref{fig8} (a).
\end{thebibliography}
\end{document}